\documentclass{article}

\usepackage[letterpaper,textheight=9in,textwidth=5.5in,top=1in]{geometry}
\usepackage[utf8]{inputenc}
\usepackage[T1]{fontenc}
\usepackage{microtype}

\usepackage{amsmath,amssymb}
\usepackage{graphicx}
\usepackage{float}
\usepackage{tikz}
\usepackage{tikz-cd}
\usetikzlibrary{arrows.meta,positioning}
\usepackage{multicol}
\usepackage[authoryear,round]{natbib}

\usepackage{xcolor}
\definecolor{linkblue}{HTML}{3B82F6}
\usepackage[breaklinks=true,colorlinks=false,citebordercolor={0 0.6 0}]{hyperref}

\makeatletter
\renewcommand{\section}{%
  \@startsection{section}{1}{\z@}%
                {-2.0ex \@plus -0.5ex \@minus -0.2ex}%
                { 1.5ex \@plus  0.3ex \@minus  0.2ex}%
                {\large\bfseries\raggedright}%
}
\renewcommand{\subsection}{%
  \@startsection{subsection}{2}{\z@}%
                {-1.8ex \@plus -0.5ex \@minus -0.2ex}%
                { 0.8ex \@plus  0.2ex}%
                {\normalsize\bfseries\raggedright}%
}
\renewcommand{\subsubsection}{%
  \@startsection{subsubsection}{3}{\z@}%
                {-1.5ex \@plus -0.5ex \@minus -0.2ex}%
                { 0.5ex \@plus  0.2ex}%
                {\normalsize\bfseries\raggedright}%
}

\newcommand{\toptitlebar}{%
  \hrule height 4pt
  \vskip 0.25in
  \vskip -\parskip%
}
\newcommand{\bottomtitlebar}{%
  \vskip 0.29in
  \vskip -\parskip%
  \hrule height 1pt
  \vskip 0.09in%
}
\renewcommand{\maketitle}{%
  \par
  \vbox{%
    \hsize\textwidth
    \linewidth\hsize
    \vskip 0.1in
    \toptitlebar
    \centering
    {\LARGE\bfseries \@title\par}
    \bottomtitlebar
    {\bfseries \@author\par}
    \ifx\@date\@empty\else
      \vspace{0.5em}{\small \@date\par}
    \fi
    \vskip 0.3in \@minus 0.1in
  }%
}
\renewenvironment{abstract}{%
  \vskip 0.075in%
  \begin{center}{\large\bfseries Abstract}\end{center}%
  \vspace{-0.5ex}%
  \begin{quote}%
}{%
  \end{quote}%
  \vskip 1ex%
}
\makeatother

\title{LLM Information Fiduciaries Improve Negotiation Outcomes}

\author{%
\normalfont
\begin{tabular}{c}
\textbf{Russell Yang}\textsuperscript{1}
\quad
\textbf{Julian Nyarko}\textsuperscript{1}\\[0.3em]
{\footnotesize \textsuperscript{1}Stanford University}\\[0.3em]
{\footnotesize
\href{mailto:jnyarko@law.stanford.edu}
{\texttt{jnyarko@law.stanford.edu}}}\\[0.1em]
\end{tabular}%
}

\begin{document}
\maketitle

\begin{abstract}
Negotiators often reach agreements that leave value on the table for both sides. We test whether a confidential LLM Information Fiduciary can recover some of this value by suggesting mutually beneficial improvements after a deal has been struck (``post-settlement settlement''). In a two-condition dyadic experiment ($N=344$ negotiating participants with $N=319$ reaching agreement), pairs negotiated the sale of a web domain. In Treatment dyads, after the negotiation concluded, the Information Fiduciary interviewed each party separately about their interests and proposed a revised proposal. This proposal replaced the original agreement only if it was accepted by both parties. In contrast, Control dyads would always stick with their original agreement. We find that participants in the Treatment dyads adopted the proposal 87\% of the time, indicating that the Information Fiduciary was able to routinely identify Pareto-dominant agreements, i.e., agreements that leave at least one party better off without harming the other side. Treatment dyads also reported higher satisfaction with their final agreement (mean difference $= 0.74$ on a 7-point scale, 95\% CI $[0.38, 1.11]$, $p < .001$, Hedges's $g = 0.44$), greater perceived fairness, and higher process and perceived counterparty satisfaction. Our results suggest that third-party LLM intermediaries can be an effective, scalable instrument that helps parties achieve better negotiation outcomes.
\end{abstract}

\section{Introduction}

Parties to negotiations often leave value on the table \citep{thompson_hastie_1990,thompson_hrebec_1996,hyder_etal_2000,bazerman_etal_2000}. Negotiation theory explains this empirical observation. Most negotiations are characterized both by distributive aspects (i.e., how the surplus from an agreement is divided) as well as integrative aspects (i.e., how much surplus the agreement generates) \citep{walton_mckersie_1965}. Integrative negotiations can require learning where the other side's priorities differ from one's own, and experiments on bilateral bargaining show that information exchange can raise joint gains while purely distributive tactics tend to depress them \citep{pruitt_lewis_1975}. A recurring obstacle to realizing integrative gains, however, is that parties do not often share their private information freely. Because the gains from disclosure are shared while the risk of exploitation falls solely on the discloser, each side withholds even though both would prefer the agreement that fuller disclosure makes possible. 

In addition to strategic concerns about exploitation, parties may fail to realize potential gains from trade because of bounded rationality \citep{simon_1955} and limited information-processing capacity. They may not fully understand their own interests, anticipate the risks and benefits implicated by the agreement, or recognize opportunities for mutually beneficial exchange. Even when the relevant interests are understood, designing an agreement that trades them off effectively can be intellectually demanding, requiring the parties to identify creative solutions to a complex optimization problem spanning numerous interdependent issues \citep{pruitt_lewis_1975,hyder_etal_2000}. The outcome is an agreement that is suboptimal for both parties compared to the alternatives that could be achieved under free information sharing and without cognitive constraints \citep{thompson_hrebec_1996}.

One way to address both challenges is to introduce a trusted expert intermediary into the negotiation. We refer to such intermediaries as ``Information Fiduciaries'': third parties to whom each side can disclose sensitive information, subject to legal or technical constraints that prevent the information from being used or disclosed to the detriment of the disclosing party \citep{balkin_2016, richards_hartzog_2016}. By receiving information that parties would be unwilling to share directly, an Information Fiduciary may be able to identify differences in their priorities and propose terms that create joint value without exposing either side to opportunistic exploitation. A sufficiently capable Information Fiduciary may also help parties clarify their own interests, identify overlooked risks and opportunities, and search over complex contractual arrangements, thereby mitigating the cognitive and computational barriers to integrative agreement.

Here, we operationalize the notion of an Information Fiduciary via large language models (LLMs) and empirically test its utility in a randomized dyadic experiment ($N=344$ negotiating participants with $N=319$ reaching agreement) over a fictional scenario involving the sale of a web domain. Control participants directly negotiated against each other to reach an agreement. In the Treatment condition, after the parties reached an initial agreement, our LLM Information Fiduciary interviewed each side confidentially and invited them to share private information. It then proposed a replacement that took effect only if both parties accepted it, an implementation of a post-settlement settlement \citep{raiffa_1985}.

Both parties accepted the Information Fiduciary's proposal in the large majority of dyads, and Treatment participants reported greater satisfaction with the final agreement, greater perceived fairness, and higher satisfaction with the negotiation process. The proposals rarely altered the negotiated price, instead adding non-price terms that participants' own agreements almost never contained. The results suggest that LLM Information Fiduciaries can serve as a scalable mechanism for producing surplus that negotiating parties fail to generate on their own.

\section{Background and Related Work}
\subsection{Uncaptured Value}

When both parties to a negotiation have complete information and are rational, unconstrained actors, writing the optimal agreement is close to a mechanical exercise. Given two parties with known utilities, the efficient deal can be computed \citep{nash_1950}, and absent any friction rational parties will continue trading until no mutually beneficial trades remain \citep{coase_1960}. This ideal ``full, open, truthful exchange'' is the ceiling that real negotiations fall short of \citep{raiffa_1982}.

Two main factors explain the gap between ideal and realized agreements. First, parties withhold information, because disclosing interests--while potentially allowing for value generation--also exposes the disclosing party to exploitation \citep{walton_mckersie_1965,lax_sebenius_1986}. What a negotiator can credibly communicate depends in part on how the other side would use the information. When the parties systematically disagree about what should be done, a negotiator may be able to convey the broad outlines of what she knows even when she cannot credibly communicate the details \citep{crawford_sobel_1982}.. Bargaining over price is an illustration that exemplifies one end of this range prone to conflict: in equilibrium both sides are incentivized to make offers that differ from their reservation prices, with buyers understating their valuations and sellers overstating theirs \citep{chatterjee_samuelson_1983}.

Second, optimizing deals is behaviorally challenging even under complete information. Negotiators often treat issues as purely distributive or zero-sum, and this conception persists throughout the negotiation process even as parties learn that issues are shared or differently valued \citep{thompson_hastie_1990,bazerman_etal_2000}. In a meta-analysis of 32 studies, parties preferred the same outcome on an issue but failed to realize it in one in five agreements \citep{thompson_hrebec_1996}. Where parties instead weigh multiple issues differently, trading across issues requires a deliberate search and optimization that can be too complex for most negotiators to undertake \citep{pruitt_lewis_1975,hyder_etal_2000}.

\subsection{Post-Settlement Settlement (PSS)}
Several procedures aim to address these challenges and help negotiators realize gains they might otherwise leave on the table. One that holds particular promise is the so-called ``post-settlement settlement'' (PSS). Introduced by \citet{raiffa_1985}, PSS first lets parties negotiate an initial agreement under the normal circumstances. After the initial agreement is reached, PSS then introduces a neutral, capable third-party Information Fiduciary. Through separate conversations, the Information Fiduciary helps parties to discover their interest and invites them to share private information without the risk of disclosure to the other side. In a last step, the Information Fiduciary proposes a new agreement that replaces the initial agreement between the parties. The new agreement will only be effective if both sides agree, ensuring that no party is worse off.

Although PSS might appear intuitively plausible, it is not immediately obvious why it should promote disclosure between rational actors. %
A pessimistic view might argue that the PSS procedure is itself just another negotiation, and so is subject to the same rational and behavioral forces that inhibit information sharing. Against this background, PSS could be viewed as a repeat of the initial negotiation, with the only difference being the parties' reservation points. Empirical evidence suggests that increasing the fallback may not improve disclosure, and may in fact hinder it  \citep{giebels_etal_2000,van_den_abbeele_etal_2009}. Further, the mere addition of a third-party mediator does not in and of itself incentivize disclosure between rational actors \citep{fey_ramsay_2010}.

However, to see how PSS could partially incentivize disclosure even among parties in conflict, consider the change of the value private information has over the course of a negotiation. During the initial bargaining process, parties often assign high value to their private information because it can be used as leverage: a Buyer who urgently needs a domain name for an upcoming product launch will conceal that deadline before the price is set, because the Seller could hold out for more. More generally, until the initial contract is formed, negotiators fail to disclose because their private information could be used by the counterparty to change the terms of the agreement in its favor. However, after the initial contract is formed, the rationale changes. Going back to the previous example, once the price is fixed, the Buyer's deadline can no longer be used by the Seller to push the price upward. More generally, after the initial agreement is formed, parties have had the opportunity to benefit from withholding the private information to gain an advantage. Because the new agreement cannot upset the initial bargain unless both parties agree, negotiators no longer have the same incentive to keep the information private. At worst, disclosure has no benefit, and the previously realized gains remain. At best, the private information can now be used to generate additional benefits for both sides. To once again tie this back to our original example, the Buyer can now reveal her deadline, and the Seller--who had avoided any sign of eagerness while the price was open--can admit that he too prefers a quick close, surfacing the shared preference at no cost to either side.

This rationale for post-settlement disclosure also has parallels in the economics of information and contract law. Here, it is commonly assumed that information has value only when it can change a decision \citep{howard_1966,hirshleifer_1971}. Perhaps closest to the present study, \citet{bar_gill_porat_2020,daughety_reinganum_2014} have applied this insight to the different stages of contracting, where an informed party might capture the value of its private information at signing before releasing it when disclosure no longer affects the terms \citep{bar_gill_porat_2020,daughety_reinganum_2014}. The most direct empirical support in favor differential disclosure across stages comes from \citet{onderstal_yang_2020}. In a procurement auction, they study a scenario in which suppliers know which project location the buyer values most and decide whether to recommend it. When that location is also the least costly for suppliers, revealing it before the auction can drive down the price through competition, whereas revealing it after the price is set allows the winning supplier to retain the cost savings. Consistent with this mechanism, the buyer's preferred location was recommended far more frequently by winning suppliers after the auction (94\%) than by bidders asked before the auction (42\%).. 

Behavioral effects on disclosure and acceptance provide additional support for PSS, especially when implemented through an AI Information Fiduciary. In personal and clinical settings, machine interviewers have been shown to reduce respondents' fear of being judged \citep{lucas_etal_2014, croes_etal_2024}. Some evidence suggests that willingness to disclose rises when respondents believe their interviewer is an AI instead of a human \citep{lucas_etal_2014}, while other work finds that disclosure to machines and humans is comparable \citep{merwin_etal_2025}. Parties also judge terms more favorably when they come from a neutral third party rather than from a counterparty \citep{ross_stillinger_1991}, which may offset the status quo bias that pushes people to keep their signed deal \citep{samuelson_zeckhauser_1988}. 

Yet despite these promising indications, PSS has rarely been empirically studied, and to the extent it has, the record appears mixed. \citet{rangaswamy_shell_1997} operationalized PSS in negotiation support software by eliciting each party's preferences in a numeric utility model, scoring tentative agreements, and generating Pareto-improving packages for consideration. They find that of fourteen dyads with an inefficient initial agreement that reached the post-settlement feature, six reopened the deal and just three improved on it. \citet{kersten_noronha_1999} took a similar approach in \textit{Inspire}, a web-based system used widely in negotiation research. Most recently, \citet{gettinger_etal_2016} studied the post-settlement phase on \textit{Inspire} at scale. However, they find that of 163 dyads that reached an inefficient agreement in a bicycle-parts scenario, 52 entered post-settlement settlement and just 21 ended with an improvement, which is about one in eight of those with a suboptimal deal.

The design of these systems may explain why productive engagement with PSS has thus far been limited. Candidate agreements were computed from the utility models stated before the negotiation, but participants reported that their preferences had shifted during the negotiation, making the suggested packages unattractive \citep{rangaswamy_shell_1997}, and \textit{Inspire} barred parties from revising their preferences at the post-settlement stage \citep{kersten_noronha_1999}. Capturing a suggested improvement also required reopening negotiation with the counterparty, which participants reported revived the uncomfortable, distributive aspects of bargaining \citep{rangaswamy_shell_1997}. Further, reopening the deal demanded additional engagement and communication between the parties. Only about half of the dyads that entered \textit{Inspire}'s post-settlement phase exchanged two or more messages, and the chance of improvement depended on the composition of the dyad, such as its gender mix and differences in the parties' language proficiency \citep{gettinger_etal_2016}. An LLM intermediary addresses these gaps: it elicits interests in ordinary conversation when preferences are current (after the initial agreement), drafts the improved package itself, and reduces the parties' effort to a brief interview and final decision.

\subsection{AI and LLMs in Negotiations}

In addition to fundamental work on negotiations, our findings contribute to an extant literature on AI and negotiations. Prior to the development of LLMs, negotiation agents were trained end-to-end on human dialogue \citep{lewis_etal_2017} and later systems separated strategic decisions from language generation \citep{he_etal_2018}. Dedicated platforms supported research on humans negotiating against virtual agents \citep{mell_gratch_2016}.

With the widespread adoption of LLMs, researchers shifted to studying language models as parties to negotiations, benchmarking them against people and against one another in controlled bargaining games \citep{kwon_etal_2024,bianchi_etal_2024,abdelnabi_etal_2024}. These studies have found LLM negotiators to be generally capable, notwithstanding some limitations. Models can perform multi-issue negotiations across many rounds, maintaining their assigned roles and inferring the other side's undisclosed priorities from text dialogue. However, they are weaker at the cooperative, value-creating reasoning required by integrative dealmaking \citep{davidson_etal_2024}, remain inferior to humans at tracking a counterpart's beliefs and intentions \citep{chan_etal_2024}, and vary in performance across models \citep{fu_etal_2023}. In LLM-human negotiations, models can be easily exploited through adversarial prompting and persistence \citep{schneider_etal_2024} and tend to bargain conservatively, conceding on trades when a human counterpart would hold out \citep{qian_etal_2026}.

Other work uses an LLM on one side but in a supporting role: an advisor who provides suggestions, a coach that performs post-hoc review of a negotiation, or a delegate who negotiates in the party's place. In a study of more than 200 human participants, participants preferred an advisor that kept them in control, although their outcomes were better under the delegate model \citep{choose_your_agent_2026}. Related simulation tools let human negotiators practice difficult conversations with a simulated counterparty \citep{mit_coaching_2025}, and assistive agents can help a party communicate in a more socially aware manner during the negotiation itself \citep{hua_etal_2024}.

Our approach differs from these studies in placing the LLM in a neutral role, meaning that the model must elicit, infer, and apply each side's interests from a short conversation. Unlike a one-sided advisor, our Information Fiduciary owes duties to both parties at once \citep{balkin_2016}. Whether fiduciary duties can workably attach to digital intermediaries is debated: the central objection is that a for-profit platform's duty of loyalty to its users is irreconcilable with its duty to shareholders, so that attaching the fiduciary label may do little beyond forestalling more structural regulation \citep{khan_pozen_2019}. The reply is that fiduciary law routinely governs actors who owe duties to multiple principals, and that the objection conflates the duties of the firm with those of its managers \citep{tuch_2021}. That objection also has less force in our setting, where the intermediary serves two individuals in a single transaction and does not monetize their information. Furthermore, because our Information Fiduciary never bargains, the weaknesses documented in bargaining settings are less material than its ability to elicit and integrate interests.

A small body of work has placed language models in this neutral capacity and provides the closest precedent for our work. An LLM that rephrased one participant's messages in a heated political exchange improved the other participant's experience without altering anyone's view \citep{argyle_etal_2023}, and LLMs given dispute-resolution problems chose interventions that raters judged as good as or better than those of human annotators \citep{tan_etal_2024}. Most directly related, the Habermas Machine gathered the private opinions of disagreeing group members and drafted consensus statements that participants preferred to those of a human mediator, moving the group toward agreement \citep{bakker_etal_2022,tessler_etal_2024}. We know of no published study that has placed a confidential language model between two real parties to propose a binding replacement that takes effect only if both accept. However, we note that, after we conducted our data collection but before the publication of this first draft, Horst Eidenmueller released a brief conceptual report similarly suggesting the theoretical opportunity offered by LLM-based fiduciaries (which he calls ``Contract Embellishers''), although without empirical testing \citep{eidenmueller_2026}.

\section{Methods}

\subsection{Study design}

We conducted a dyadic experiment using a two-group, posttest-only design (exempt human-subjects research, Protocol ID 83436). The experiment was conducted in January 2026. All participants provided informed consent before beginning the study.

Participants negotiated a modified version of the Harvard Law School Program on Negotiation's \textit{Lattitude.com} case \citep{manwaring_lattitude_2003}. The scenario concerned the sale of the fictional web domain from a geography teacher, who had mistakenly purchased it, to a small business owner.

We recruited adult participants based in the United States through Prolific, which randomly allocated each participant to one of two arms; within each condition, participants were paired into two-person dyads. In each dyad, participants navigated through a customized environment to complete their negotiation.\footnote{We used the platform Labvanced to design the environment.} One participant was assigned to the role of the buyer and the other to the role of the seller of the domain. Parties negotiated through a text-chat interface that also allowed them to make binding offers (Appendix~\ref{app:procedure}). An exit survey was administered after the negotiation procedure was complete and the final agreement had been determined (Appendix~\ref{app:survey}).

The study instructions (see Appendix~\ref{app:instructions}) stated that participants would receive a base compensation and a performance-based bonus, with performance being dependent on the benefits the final agreement generated for the participants (both pecuniary and non-pecuniary). To enable integrative bargaining solutions, the participants were specifically instructed that they do not compete with the other side for the bonus, and that they should only focus on the benefits generated for themselves. Thus, both members of a dyad could independently receive high bonuses if the final agreement was good for both sides. The study instructions also stated that survey responses would not be used to calculate the bonus.

\subsection{Experimental conditions}

In the Control condition, dyads negotiated with each other using a text-chat and binding-offer interface, and any agreement reached was their final agreement. The Treatment condition added the post-negotiation LLM Information Fiduciary intervention after the initial negotiation, where Treatment dyads that reached an initial agreement then interacted with the Information Fiduciary (GPT-4.1) in two stages.

In the first stage, an information-collector agent privately interviewed each party with the goal of identifying their private interests. The agent received  the initial agreement and the negotiation transcript, but had no access to a party's confidential role instructions. Parties were free to volunteer information as they saw fit, and could copy and paste from their instructions for convenience. The agent was instructed to open with an invitation to share background information before proposing potential terms or areas of improvement that might be of interest to the party.

In the second stage, a proposal-generator agent received the initial agreement together with both interview transcripts and synthesized them into a new proposal intended to improve the agreement for both parties.
Each party independently chose whether to accept or reject the new proposal, which replaced the initial agreement only if both parties accepted it.
\subsection{Outcomes}

The primary outcome was participants' satisfaction with the final agreement, measured on a seven-point Likert scale after the final agreement was determined. The secondary outcomes were participants' perceptions of the other side's satisfaction, the fairness of the agreement to both sides, the clarity and implementability of the agreement, and satisfaction with the negotiation process. These measures reflect established dimensions of subjective value in negotiation: the instrumental outcome, the negotiation process, and the relationship with the counterparty \citep{curhan_elfenbein_xu_2006}.

We also analyzed agreement content at the dyad level among dyads that reached agreement. These analyses described whether treatment dyads adopted the LLM-generated proposal, how the proposed monetary terms compared with the initial agreement, and what kinds of non-price terms appeared in the initial and proposed agreements.

\subsection{Statistical analysis}

We conducted statistical analysis in R (version 4.5.1). For each survey outcome, we estimated the treatment effect as the difference in condition means, $\Delta = \bar{Y}_{T} - \bar{Y}_{C}$. The primary outcome was analyzed using a two-sided Welch two-sample t-test. We also report condition means, $\Delta$ with its 95\% confidence interval, the two-sided $p$ value, and Hedges's $g$. We applied the same approach to secondary outcomes, with $p$ values adjusted using the Benjamini--Hochberg procedure to account for multiple comparisons across the four secondary outcomes.

For treatment dyads, we determined whether the LLM-generated proposal was accepted by both parties, rejected by both parties, or accepted by only one party. In addition, we analyzed monetary proposals, changes from the initial agreement, and the prevalence and categorization of non-price agreement terms. Non-price terms were extracted verbatim from agreement text and assigned to categories using an LLM (GPT-5.5, reasoning effort ``high''). The category set was LLM-discovered subject to three user-specified categories (Redundant Terms, Bespoke, and Nonsensical). Dyad-level content analyses were restricted to pairs in which both members completed the study.\footnote{In three pairs (two Control, one Treatment), the negotiation concluded with an agreement, but one participant dropped out before completing the rest of the study. These pairs are excluded from the dyad-level content analyses, but those who completed it are still included in the participant-level survey analyses.}

As an exploratory analysis, we used the same LLM (GPT 5.5, reasoning effort ``high'') to code the information-collector interview transcripts from the 76 Treatment negotiations that reached agreement (one buyer and one seller interview per negotiation, including the singleton group; 152 interviews) into preference items, recording for each interview and item whether the participant expressed wanting it (Yes), not wanting it (No), a mixed or conditional stance, or no clear stance, and ranked items by endorsement rate (Yes divided by coded responses, which exclude unclear and not-mentioned items). We also compared proposal acceptance by participant gender using Fisher's exact test.

Because members of a dyad negotiated with each other, their responses may be correlated. As a robustness check, we re-estimated each survey-outcome treatment effect as an OLS regression of the outcome on a treatment indicator with cluster-robust standard errors clustered at the dyad level and Satterthwaite degrees of freedom (Appendix Table~\ref{tab:robustness}).

\section{Results}

\begin{figure}[H]
\centering
\begin{tikzpicture}[
  mainbox/.style={draw, rounded corners=1.5pt, align=center, text width=2.85cm, minimum height=0.88cm, inner sep=4pt, font=\footnotesize},
  sidebox/.style={draw=gray!65, fill=gray!8, rounded corners=1.5pt, align=center, text width=2.35cm, minimum height=0.88cm, inner sep=4pt, font=\footnotesize},
  flowhead/.style={font=\bfseries\small},
  flowarrow/.style={-{Stealth[length=2.4mm]}, line width=0.45pt},
  sidearrow/.style={-{Stealth[length=2.2mm]}, line width=0.35pt, draw=gray!65}
]

\node[mainbox] (t0) at (0,0) {Prolific submissions\\$n=200$};
\node[flowhead, above=0.28cm of t0] {Treatment};
\node[mainbox, below=0.75cm of t0] (t1) {Completed on Prolific\\$n=166$};
\node[sidebox, right=0.65cm of t1] (t1x) {Not Completed\\$n=34$};
\node[mainbox, below=0.75cm of t1] (t2) {Labvanced records\\$n=166$};
\node[mainbox, below=0.75cm of t2] (t3) {Agreement\\$n=151$};
\node[sidebox, right=0.65cm of t3] (t3x) {No agreement\\$n=15$};
\node[mainbox, below=0.75cm of t3] (t6) {Two-person dyad survey finishers\\$n=150$ (75 dyads)};
\node[sidebox, right=0.65cm of t6] (t6x) {Singleton survey\\finishers\\$n=1$};

\draw[flowarrow] (t0.south) -- coordinate[midway] (tcomplete) (t1.north);
\draw[sidearrow] (tcomplete) -| (t1x.north);
\draw[flowarrow] (t1.south) -- (t2.north);
\draw[flowarrow] (t2.south) -- coordinate[midway] (tagreement) (t3.north);
\draw[sidearrow] (tagreement) -| (t3x.north);
\draw[flowarrow] (t3.south) -- coordinate[midway] (tsplit) (t6.north);
\draw[sidearrow] (tsplit) -| (t6x.north);

\node[mainbox] (c0) at (7.15,0) {Prolific submissions\\$n=199$};
\node[flowhead, above=0.28cm of c0] {Control};
\node[mainbox, below=0.75cm of c0] (c1) {Completed on Prolific\\$n=178$};
\node[sidebox, right=0.65cm of c1] (c1x) {Not Completed\\$n=21$};
\node[mainbox, below=0.75cm of c1] (c2) {Labvanced records\\$n=178$};
\node[mainbox, below=0.75cm of c2] (c3) {Agreement\\$n=168$};
\node[sidebox, right=0.65cm of c3] (c3x) {No agreement\\$n=10$};
\node[mainbox, below=0.75cm of c3] (c5) {Two-person dyad survey finishers\\$n=166$ (83 dyads)};
\node[sidebox, right=0.65cm of c5] (c5x) {Singleton survey\\finishers\\$n=2$};

\draw[flowarrow] (c0.south) -- coordinate[midway] (ccomplete) (c1.north);
\draw[sidearrow] (ccomplete) -| (c1x.north);
\draw[flowarrow] (c1.south) -- (c2.north);
\draw[flowarrow] (c2.south) -- coordinate[midway] (cagreement) (c3.north);
\draw[sidearrow] (cagreement) -| (c3x.north);
\draw[flowarrow] (c3.south) -- coordinate[midway] (csplit) (c5.north);
\draw[sidearrow] (csplit) -| (c5x.north);
\draw[gray!25, line width=0.35pt] (5.35,0.75) -- (5.35,-7.8);
\end{tikzpicture}
\caption{Participant flow by experimental condition. The primary survey analysis included the 319 participants who reached agreement and reported their satisfaction with the final agreement. Dyad-level agreement-content analyses exclude singletons.}
\label{fig:participant-flow}
\end{figure}
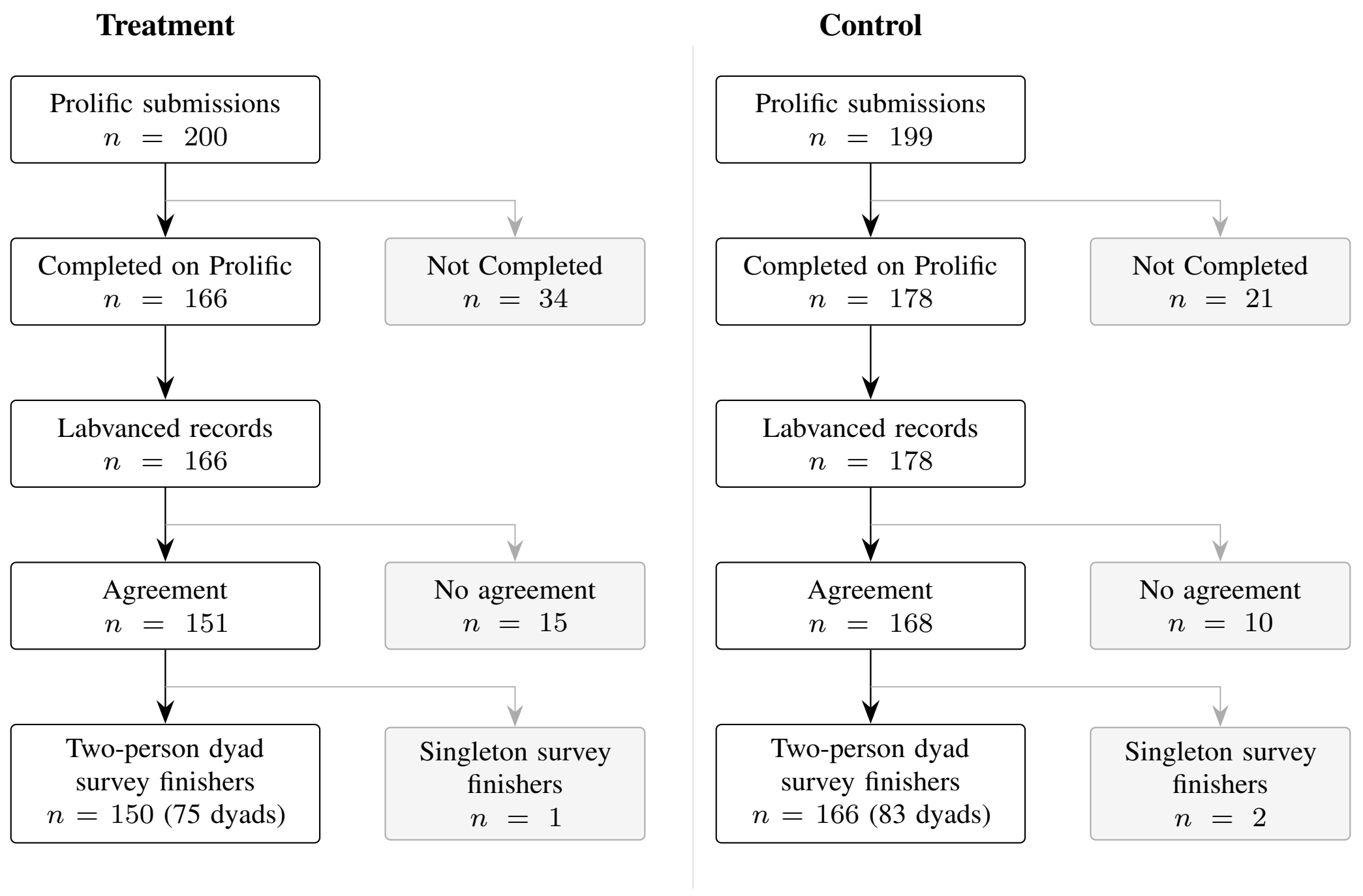

Participant flow is shown in Figure~\ref{fig:participant-flow}.

\subsection{Survey outcomes}

\begin{figure}[H]
\centering
\includegraphics[width=0.94\textwidth]{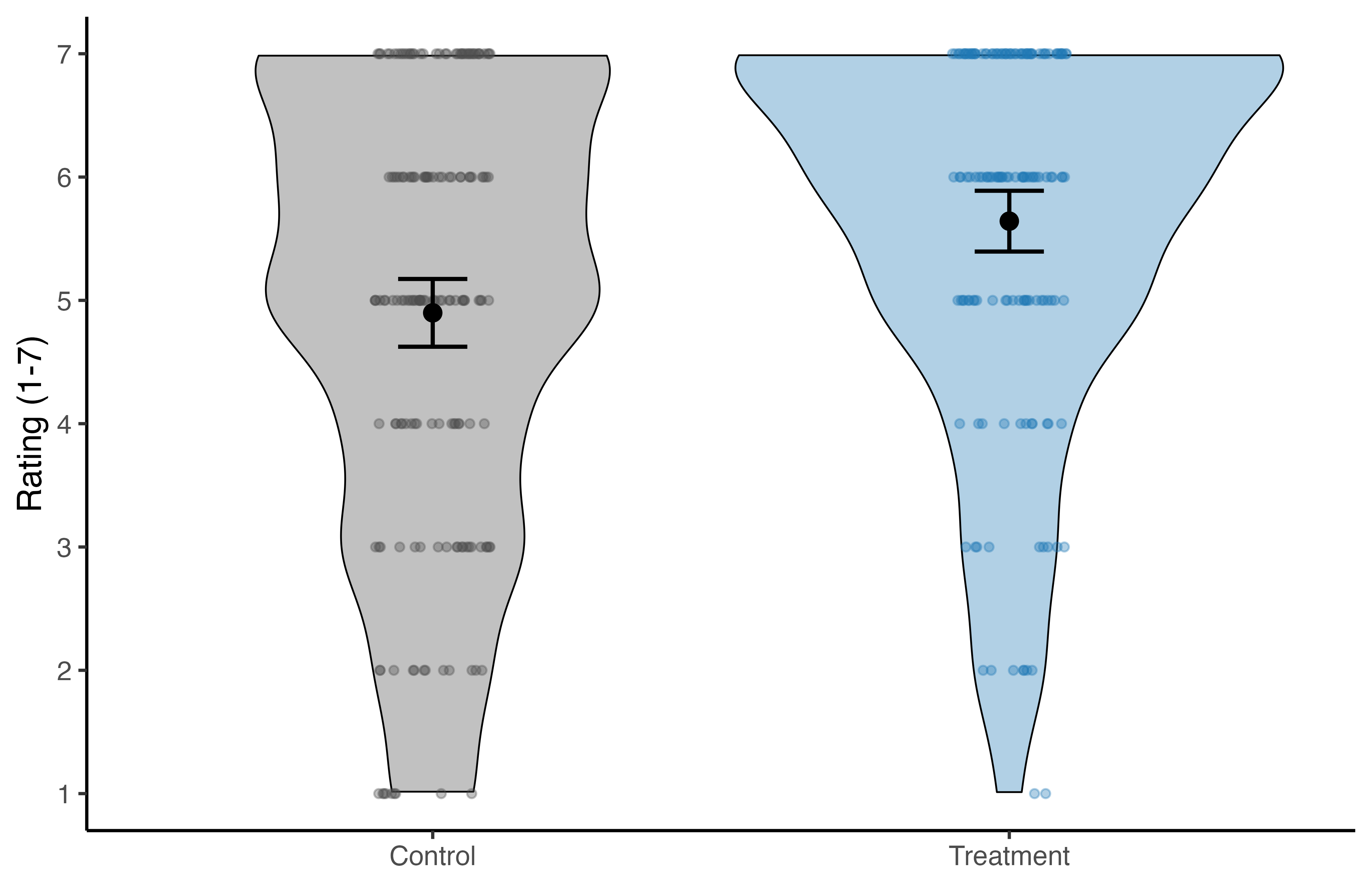}
\caption{Distribution of the primary outcome by condition. Violin plots show the distribution of Likert ratings with individual responses jittered; points and error bars indicate means and 95\% confidence intervals.}
\label{fig:primary}
\end{figure}

Survey outcomes were analyzed at the participant level among those who completed the survey (Treatment $n=151$, Control $n=168$). Satisfaction with the final agreement was higher in the Treatment condition than in the Control condition ($\Delta=0.74$ on the 7-point scale, 95\% CI $[0.38, 1.11]$, $t(316.0)=3.98$, $p<.001$, Hedges's $g=0.44$; Table~\ref{tab:primary}, Figure~\ref{fig:primary}).

\begin{figure}[H]
\centering
\includegraphics[width=0.94\textwidth]{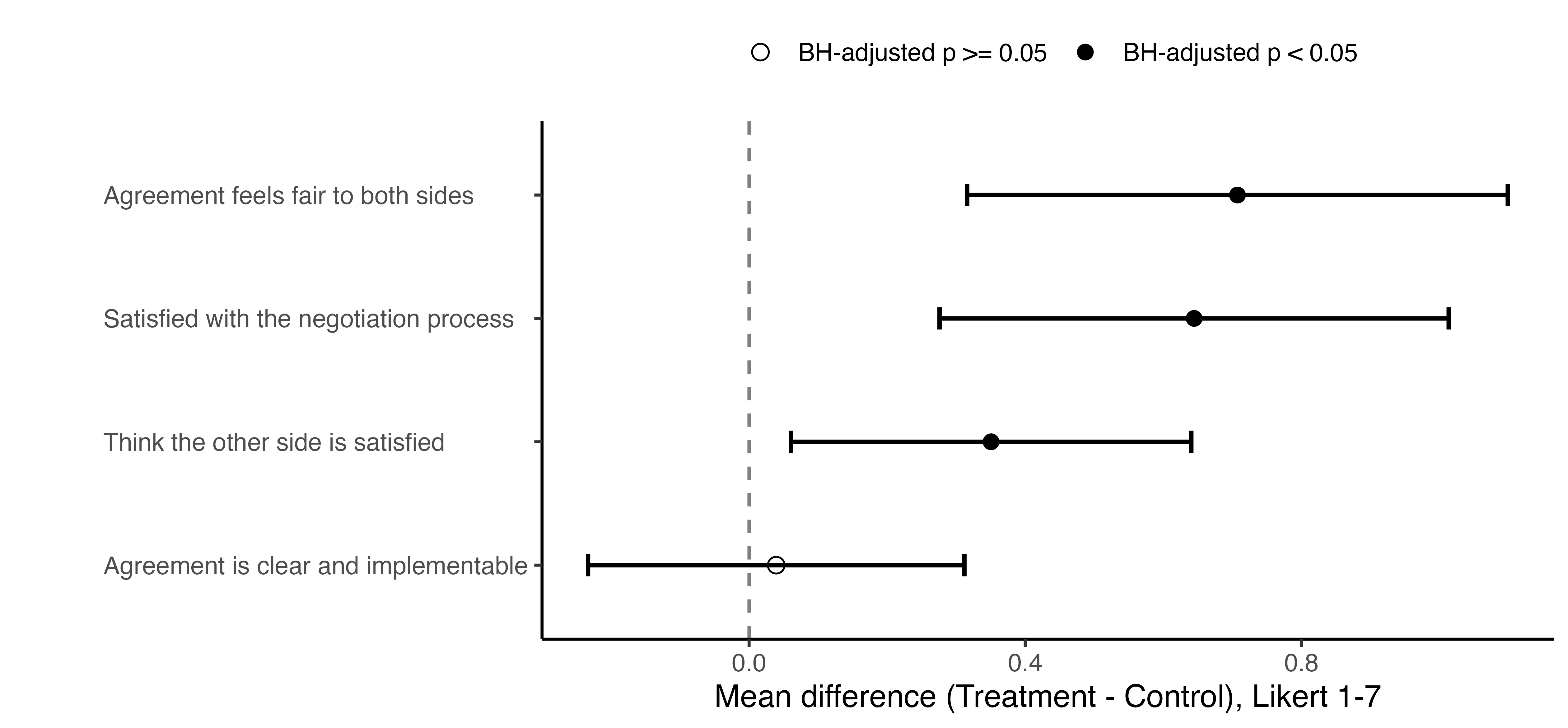}
\caption{Secondary outcomes: mean difference (Treatment $-$ Control) with 95\% confidence intervals. Filled points indicate BH-adjusted $p<.05$.}
\label{fig:secondary}
\end{figure}

Secondary outcomes showed a similar pattern (Table~\ref{tab:secondary}, Figure~\ref{fig:secondary}). After Benjamini--Hochberg correction, Treatment participants reported higher perceived fairness of the agreement ($g=0.40$), higher satisfaction with the negotiation process ($g=0.38$), and higher perceived satisfaction of the other side ($g=0.26$; all BH-adjusted $p<.05$). The clarity and implementability of the agreement did not differ significantly between conditions. Appendix Figure~\ref{fig:distributions} shows the response distribution for each outcome by condition.

Estimates were essentially unchanged under dyad-clustered standard errors (Appendix Table~\ref{tab:robustness}).

\begin{table}[H]
\centering
\caption{Primary survey outcome.}
\label{tab:primary}
\resizebox{\textwidth}{!}{%
\begin{tabular}{lcccccc}
\hline
Outcome & Control mean (SD) & Treatment mean (SD) & Mean difference [95\% CI] & $t(df)$ & $p$ & Hedges's $g$ \\
\hline
Satisfied with the final agreement & 4.90 (1.80) & 5.64 (1.53) & 0.74 [0.38, 1.11] & 3.98 (316.0) & $<.001$ & 0.44 \\
\hline
\end{tabular}%
}
\end{table}

\begin{table}[H]
\centering
\caption{Secondary survey outcomes with Benjamini--Hochberg correction.}
\label{tab:secondary}
\resizebox{\textwidth}{!}{%
\begin{tabular}{lccccc}
\hline
Outcome & Control mean (SD) & Treatment mean (SD) & Mean difference [95\% CI] & BH-adjusted $p$ & Hedges's $g$ \\
\hline
Agreement feels fair to both sides & 4.84 (1.82) & 5.55 (1.72) & 0.71 [0.32, 1.10] & .001 & 0.40 \\
Satisfied with the negotiation process & 5.20 (1.81) & 5.84 (1.54) & 0.64 [0.28, 1.01] & .001 & 0.38 \\
Think the other side is satisfied & 5.64 (1.43) & 5.99 (1.20) & 0.35 [0.06, 0.64] & .024 & 0.26 \\
Agreement is clear and implementable & 6.17 (1.35) & 6.21 (1.12) & 0.04 [-0.23, 0.31] & .777 & 0.03 \\
\hline
\end{tabular}%
}
\end{table}

\subsection{Agreement content}

\begin{figure}[H]
\centering
\includegraphics[width=0.86\textwidth]{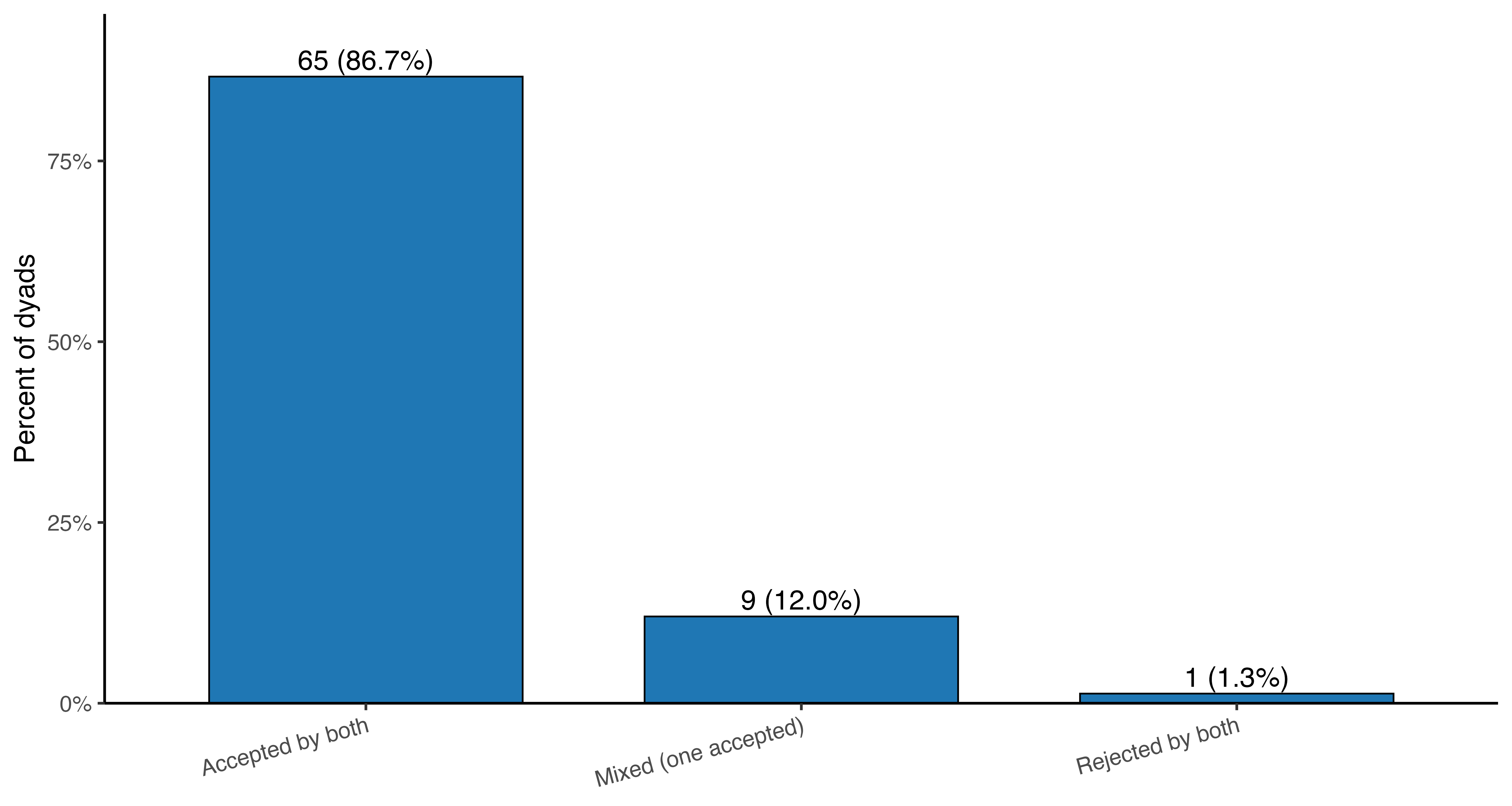}
\caption{New proposal acceptance among Treatment dyads that reached agreement. Bars show the percentage of dyads where both parties accepted, acceptance was mixed, or both parties rejected.}
\label{fig:acceptance}
\end{figure}

Descriptive analyses of agreement content were performed at the dyad level among dyads that reached agreement. Among the 75 Treatment dyads that reached an initial agreement, the LLM-generated proposal was accepted by both parties in 65 dyads (86.7\%), accepted by exactly one party in 9 dyads (12.0\%), and rejected by both parties in 1 dyad (1.3\%; Figure~\ref{fig:acceptance}). Acceptance did not differ by gender (Appendix~\ref{app:gender}).

The proposed monetary price matched the initial agreement price in 73 of the 75 dyads; it was lower in the proposal in 2 dyads and higher in none (Appendix Figure~\ref{fig:price}). The median price was \$2{,}500 in both the initial agreements and the proposals.

Non-price (``other'') terms were infrequent in initial agreements and present in every proposal (Appendix Figure~\ref{fig:presence}). The ``other terms'' field contained at least one clause in 8 of 83 Control initial agreements (9.6\%), 12 of 75 Treatment initial agreements (16.0\%), and in all 75 LLM proposals (100\%).

The non-price terms in the proposals spanned several substantive categories (Appendix Figure~\ref{fig:categories}). Proposals specified closing timing and payment terms in 70 of 75 dyads, transfer logistics and fees in 65, ancillary covenants in 44, domain assurances and assets in 33, and written documentation in 21. Non-price terms in the initial agreements were sparse and consisted mostly of restatements of price or ownership (13 dyads).

Appendix Table~\ref{tab:preferences} shows the LLM-coded preference items derived from the information-collector interviews. Public recognition or promotion was the most frequently mentioned item (109 of 152 interviews), followed by transfer responsibility and assistance (95), payment structure (83), and transfer timing and closing deadline (81). Among coded responses, endorsement rates ranged from 22.8\% to 100\%, with the highest for clean reputation and technical assurances (7 of 7) and ownership and legal assurances (16 of 18).

\section{Discussion}

A confidential LLM Information Fiduciary, added after dyads had already reached an agreement, raised satisfaction with the final agreement ($\Delta=0.74$, 95\% CI $[0.38, 1.11]$, $t(316.0)=3.98$, $p<.001$, $g=0.44$), along with perceived fairness ($g=0.40$), process satisfaction ($g=0.38$), and perceived counterparty satisfaction ($g=0.26$; all BH-adjusted $p<.05$). The content analyses show that proposals almost never moved the negotiated price (73 of 75 dyads) but added non-price terms (payment timing and method, transfer logistics and fees, assurances, documentation) that were largely absent from the dyads' own agreements (present in 9.6\% of Control and 16.0\% of Treatment initial agreements, versus 100\% of proposals), and both parties accepted the new LLM proposal in 86.7\% of dyads.

The results provide an experimental demonstration of post-settlement settlement \citep{raiffa_1985} with real human dyads, with an LLM performing the search role that earlier negotiation support systems implemented through structured preference elicitation \citep{rangaswamy_shell_1997}. Consistent with evidence that negotiators fail to capture integrative value \citep{thompson_hrebec_1996}, the dyads' own agreements rarely contained non-price terms, and a confidential intermediary captured part of that value without requiring disclosure to the other side \citep{ehtamo_etal_1999}. The near-universal acceptance rate of LLM-generated proposals suggests that participants treated the system as a neutral intermediary rather than another strategic actor, an instantiation by design of the fiduciary-like duties debated in the legal literature \citep{balkin_2016,khan_pozen_2019, tuch_2021}.

\section{Limitations}
We ran a single stylized scenario with modest stakes, short online negotiations, and immediate self-reported outcomes rather than realized payoffs. Although participants were randomly allocated to arms, the arms filled at different rates after their same-day launch, so completion times were not perfectly balanced across conditions. In addition, because Control dyads experienced no comparable post-negotiation interaction, part of the satisfaction effect may reflect the added attention of the interview, demand effects, or the framing of receiving a ``win-win'' offer rather than the substance of the proposed terms. Future work could use objectively scoreable multi-issue payoffs, include an active control for the post-negotiation interaction, or test whether proposed terms are implemented and valued after the fact.

\bibliographystyle{plainnat}
\bibliography{references}

@misc{manwaring_lattitude_2003,
  author       = {Manwaring, Melissa},
  title        = {Lattitude.com},
  year         = {2003},
  howpublished = {Negotiation exercise (teaching material), Program on Negotiation at Harvard Law School, Teaching Negotiation Resource Center},
  note         = {Case no.\ PON544}
}

@article{nash_1950,
  title   = {The Bargaining Problem},
  author  = {Nash, John F.},
  journal = {Econometrica},
  volume  = {18},
  number  = {2},
  pages   = {155--162},
  year    = {1950}
}

@article{coase_1960,
  title   = {The Problem of Social Cost},
  author  = {Coase, Ronald H.},
  journal = {Journal of Law and Economics},
  volume  = {3},
  pages   = {1--44},
  year    = {1960}
}

@book{raiffa_1982,
  title     = {The Art and Science of Negotiation},
  author    = {Raiffa, Howard},
  publisher = {Harvard University Press},
  address   = {Cambridge, MA},
  year      = {1982}
}

@book{lax_sebenius_1986,
  title     = {The Manager as Negotiator: Bargaining for Cooperation and Competitive Gain},
  author    = {Lax, David A. and Sebenius, James K.},
  publisher = {Free Press},
  address   = {New York},
  year      = {1986}
}

@article{chatterjee_samuelson_1983,
  title   = {Bargaining under Incomplete Information},
  author  = {Chatterjee, Kalyan and Samuelson, William},
  journal = {Operations Research},
  volume  = {31},
  number  = {5},
  pages   = {835--851},
  year    = {1983},
  doi     = {10.1287/opre.31.5.835}
}

@article{crawford_sobel_1982,
  title   = {Strategic Information Transmission},
  author  = {Crawford, Vincent P. and Sobel, Joel},
  journal = {Econometrica},
  volume  = {50},
  number  = {6},
  pages   = {1431--1451},
  year    = {1982}
}

@article{simon_1955,
  title   = {A Behavioral Model of Rational Choice},
  author  = {Simon, Herbert A.},
  journal = {Quarterly Journal of Economics},
  volume  = {69},
  number  = {1},
  pages   = {99--118},
  year    = {1955},
  doi     = {10.2307/1884852}
}

@article{kersten_noronha_1999,
  title   = {{WWW}-Based Negotiation Support: Design, Implementation, and Use},
  author  = {Kersten, Gregory E. and Noronha, Sunil J.},
  journal = {Decision Support Systems},
  volume  = {25},
  number  = {2},
  pages   = {135--154},
  year    = {1999},
  doi     = {10.1016/S0167-9236(99)00012-3}
}

@article{gettinger_etal_2016,
  title   = {Why Can't We Settle Again? Analysis of Factors That Influence Agreement Prospects in the Post-Settlement Phase},
  author  = {Gettinger, Johannes and Filzmoser, Michael and Koeszegi, Sabine T.},
  journal = {Journal of Business Economics},
  volume  = {86},
  number  = {4},
  pages   = {413--440},
  year    = {2016},
  doi     = {10.1007/s11573-016-0809-5}
}

@article{fey_ramsay_2010,
  title   = {When Is Shuttle Diplomacy Worth the Commute? Information Sharing through Mediation},
  author  = {Fey, Mark and Ramsay, Kristopher W.},
  journal = {World Politics},
  volume  = {62},
  number  = {4},
  pages   = {529--560},
  year    = {2010},
  doi     = {10.1017/S0043887110000183}
}

@article{howard_1966,
  title   = {Information Value Theory},
  author  = {Howard, Ronald A.},
  journal = {IEEE Transactions on Systems Science and Cybernetics},
  volume  = {2},
  number  = {1},
  pages   = {22--26},
  year    = {1966},
  doi     = {10.1109/TSSC.1966.300074}
}

@article{hirshleifer_1971,
  title   = {The Private and Social Value of Information and the Reward to Inventive Activity},
  author  = {Hirshleifer, Jack},
  journal = {American Economic Review},
  volume  = {61},
  number  = {4},
  pages   = {561--574},
  year    = {1971}
}

@article{bar_gill_porat_2020,
  title   = {Disclosure Rules in Contract Law},
  author  = {Bar-Gill, Oren and Porat, Ariel},
  journal = {Journal of Legal Studies},
  volume  = {49},
  number  = {1},
  pages   = {103--152},
  year    = {2020},
  doi     = {10.1086/707996}
}

@article{daughety_reinganum_2014,
  title   = {Revelation and Suppression of Private Information in Settlement-Bargaining Models},
  author  = {Daughety, Andrew F. and Reinganum, Jennifer F.},
  journal = {University of Chicago Law Review},
  volume  = {81},
  number  = {1},
  pages   = {83--108},
  year    = {2014}
}

@techreport{onderstal_yang_2020,
  title       = {Cheap-Talk Communication in Procurement Auctions: Theory and Experiment},
  author      = {Onderstal, Sander and Yang, Yang},
  institution = {Tinbergen Institute},
  number      = {Discussion Paper TI 2020-013/VII},
  year        = {2020},
  url         = {https://papers.tinbergen.nl/20013.pdf}
}

@article{giebels_etal_2000,
  title   = {Interdependence in Negotiation: Effects of Exit Options and Social Motive on Distributive and Integrative Negotiation},
  author  = {Giebels, Ellen and De Dreu, Carsten K. W. and Van de Vliert, Evert},
  journal = {European Journal of Social Psychology},
  volume  = {30},
  number  = {2},
  pages   = {255--272},
  year    = {2000}
}

@article{van_den_abbeele_etal_2009,
  title   = {The Effect of Cost Information on Buyer-Supplier Negotiations in Different Power Settings},
  author  = {Van den Abbeele, Alexandra and Roodhooft, Filip and Warlop, Luk},
  journal = {Accounting, Organizations and Society},
  volume  = {34},
  number  = {2},
  pages   = {245--266},
  year    = {2009},
  doi     = {10.1016/j.aos.2008.05.005}
}

@article{samuelson_zeckhauser_1988,
  title   = {Status Quo Bias in Decision Making},
  author  = {Samuelson, William and Zeckhauser, Richard},
  journal = {Journal of Risk and Uncertainty},
  volume  = {1},
  number  = {1},
  pages   = {7--59},
  year    = {1988},
  doi     = {10.1007/BF00055564}
}

@article{ross_stillinger_1991,
  title   = {Barriers to Conflict Resolution},
  author  = {Ross, Lee and Stillinger, Constance},
  journal = {Negotiation Journal},
  volume  = {7},
  number  = {4},
  pages   = {389--404},
  year    = {1991},
  doi     = {10.1111/j.1571-9979.1991.tb00634.x}
}

@article{lucas_etal_2014,
  title   = {It's Only a Computer: Virtual Humans Increase Willingness to Disclose},
  author  = {Lucas, Gale M. and Gratch, Jonathan and King, Aisha and Morency, Louis-Philippe},
  journal = {Computers in Human Behavior},
  volume  = {37},
  pages   = {94--100},
  year    = {2014},
  doi     = {10.1016/j.chb.2014.04.043}
}

@article{croes_etal_2024,
  title   = {Digital Confessions: The Willingness to Disclose Intimate Information to a Chatbot and Its Impact on Emotional Well-Being},
  author  = {Croes, Emmelyn A. J. and Antheunis, Marjolijn L. and van der Lee, Chris and de Wit, Jan M. S.},
  journal = {Interacting with Computers},
  volume  = {36},
  number  = {5},
  pages   = {279--292},
  year    = {2024}
}

@article{merwin_etal_2025,
  title   = {Self-Disclosure to {AI}: People Provide Personal Information to {AI} and Humans Equivalently},
  author  = {Merwin, E. R. and Hagen, A. C. and Keebler, J. R. and Forbes, C.},
  journal = {Computers in Human Behavior: Artificial Humans},
  volume  = {5},
  pages   = {100180},
  year    = {2025},
  doi     = {10.1016/j.chbah.2025.100180}
}

@article{curhan_elfenbein_xu_2006,
  title   = {What Do People Value When They Negotiate? Mapping the Domain of Subjective Value in Negotiation},
  author  = {Curhan, Jared R. and Elfenbein, Hillary Anger and Xu, Heng},
  journal = {Journal of Personality and Social Psychology},
  volume  = {91},
  number  = {3},
  pages   = {493--512},
  year    = {2006},
  doi     = {10.1037/0022-3514.91.3.493}
}

@inproceedings{lewis_etal_2017,
  title     = {Deal or No Deal? End-to-End Learning of Negotiation Dialogues},
  author    = {Lewis, Mike and Yarats, Denis and Dauphin, Yann N. and Parikh, Devi and Batra, Dhruv},
  booktitle = {Proceedings of the 2017 Conference on Empirical Methods in Natural Language Processing (EMNLP)},
  pages     = {2443--2453},
  year      = {2017}
}

@inproceedings{mell_gratch_2016,
  title     = {{IAGO}: Interactive Arbitration Guide Online (Demonstration)},
  author    = {Mell, Johnathan and Gratch, Jonathan},
  booktitle = {Proceedings of the 2016 International Conference on Autonomous Agents and Multiagent Systems (AAMAS)},
  pages     = {1510--1512},
  year      = {2016}
}

@inproceedings{davidson_etal_2024,
  title     = {Evaluating Language Model Agency through Negotiations},
  author    = {Davidson, Tim R. and Veselovsky, Veniamin and Josifoski, Martin and Peyrard, Maxime and Bosselut, Antoine and Kosinski, Michal and West, Robert},
  booktitle = {International Conference on Learning Representations (ICLR)},
  year      = {2024}
}

@inproceedings{chan_etal_2024,
  title     = {{NegotiationToM}: A Benchmark for Stress-Testing Machine Theory of Mind on Negotiation Surrounding},
  author    = {Chan, Chunkit and Jiayang, Cheng and Yim, Yauwai and Deng, Zheye and Fan, Wei and Li, Haoran and Liu, Xin and Zhang, Hongming and Wang, Weiqi and Song, Yangqiu},
  booktitle = {Findings of the Association for Computational Linguistics: EMNLP 2024},
  pages     = {4211--4241},
  year      = {2024}
}

@article{argyle_etal_2023,
  title   = {Leveraging {AI} for Democratic Discourse: Chat Interventions Can Improve Online Political Conversations at Scale},
  author  = {Argyle, Lisa P. and Bail, Christopher A. and Busby, Ethan C. and Gubler, Joshua R. and Howe, Thomas and Rytting, Christopher and Sorensen, Taylor and Wingate, David},
  journal = {Proceedings of the National Academy of Sciences},
  volume  = {120},
  number  = {41},
  pages   = {e2311627120},
  year    = {2023},
  doi     = {10.1073/pnas.2311627120}
}

@inproceedings{tan_etal_2024,
  title     = {Robots in the Middle: Evaluating LLMs in Dispute Resolution},
  author    = {Tan, Jinzhe and Westermann, Hannes and Pottanigari, Nikhil Reddy and {\v{S}}avelka, Jarom{\'i}r and Mee{\`u}s, S{\'e}bastien and Godet, Mia and Benyekhlef, Karim},
  booktitle = {Legal Knowledge and Information Systems (JURIX 2024)},
  pages     = {168--179},
  publisher = {IOS Press},
year      = {2024},
  doi       = {10.3233/FAIA241243}
}

@inproceedings{bakker_etal_2022,
  title     = {Fine-Tuning Language Models to Find Agreement among Humans with Diverse Preferences},
  author    = {Bakker, Michiel A. and Chadwick, Martin J. and Sheahan, Hannah R. and Tessler, Michael Henry and Campbell-Gillingham, Lucy and Balaguer, Jan and McAleese, Nat and Glaese, Amelia and Aslanides, John and Botvinick, Matthew M. and Summerfield, Christopher},
  booktitle = {Advances in Neural Information Processing Systems (NeurIPS)},
  volume    = {35},
  year      = {2022}
}

@article{tessler_etal_2024,
  title   = {{AI} Can Help Humans Find Common Ground in Democratic Deliberation},
  author  = {Tessler, Michael Henry and Bakker, Michiel A. and Jarrett, Daniel and Sheahan, Hannah and Chadwick, Martin J. and Koster, Raphael and Evans, Georgina and Campbell-Gillingham, Lucy and Collins, Tantum and Parkes, David C. and Botvinick, Matthew and Summerfield, Christopher},
  journal = {Science},
  volume  = {386},
  number  = {6719},
  pages   = {eadq2852},
  year    = {2024},
  doi     = {10.1126/science.adq2852}
}

@article{thompson_hastie_1990,
  title     = {Social Perception in Negotiation},
  author    = {Thompson, Leigh and Hastie, Reid},
  journal   = {Organizational Behavior and Human Decision Processes},
  volume    = {47},
  number    = {1},
  pages     = {98--123},
  year      = {1990},
  publisher = {Elsevier},
  doi       = {10.1016/0749-5978(90)90048-E}
}

@article{hyder_etal_2000,
  title   = {Getting to Best: Efficiency versus Optimality in Negotiation},
  author  = {Hyder, Elaine B. and Prietula, Michael J. and Weingart, Laurie R.},
  journal = {Cognitive Science},
  volume  = {24},
  number  = {2},
  pages   = {169--204},
  year    = {2000},
  doi     = {10.1207/s15516709cog2402_1}
}

@article{bazerman_etal_2000,
  title   = {Negotiation},
  author  = {Bazerman, Max H. and Curhan, Jared R. and Moore, Don A. and Valley, Kathleen L.},
  journal = {Annual Review of Psychology},
  volume  = {51},
  pages   = {279--314},
  year    = {2000},
  doi     = {10.1146/annurev.psych.51.1.279}
}

@book{walton_mckersie_1965,
  title     = {A Behavioral Theory of Labor Negotiations: An Analysis of a Social Interaction System},
  author    = {Walton, Richard E. and McKersie, Robert B.},
  publisher = {McGraw-Hill},
  address   = {New York},
  year      = {1965}
}

@article{pruitt_lewis_1975,
  title   = {Development of Integrative Solutions in Bilateral Negotiation},
  author  = {Pruitt, Dean G. and Lewis, Steven A.},
  journal = {Journal of Personality and Social Psychology},
  volume  = {31},
  number  = {4},
  pages   = {621--633},
  year    = {1975},
  doi     = {10.1037/0022-3514.31.4.621}
}

@article{thompson_hrebec_1996,
  title   = {Lose-Lose Agreements in Interdependent Decision Making},
  author  = {Thompson, Leigh and Hrebec, Dennis},
  journal = {Psychological Bulletin},
  volume  = {120},
  number  = {3},
  pages   = {396--409},
  year    = {1996},
  doi     = {10.1037/0033-2909.120.3.396}
}

@article{balkin_2016,
  title   = {Information Fiduciaries and the First Amendment},
  author  = {Balkin, Jack M.},
  journal = {UC Davis Law Review},
  volume  = {49},
  number  = {4},
  pages   = {1183--1234},
  year    = {2016}
}

@article{richards_hartzog_2016,
  title   = {Taking Trust Seriously in Privacy Law},
  author  = {Richards, Neil M. and Hartzog, Woodrow},
  journal = {Stanford Technology Law Review},
  volume  = {19},
  pages   = {431--472},
  year    = {2016}
}

@article{khan_pozen_2019,
  title   = {A Skeptical View of Information Fiduciaries},
  author  = {Khan, Lina M. and Pozen, David E.},
  journal = {Harvard Law Review},
  volume  = {133},
  number  = {2},
  pages   = {497--541},
  year    = {2019}
}

@article{tuch_2021,
  title   = {A General Defense of Information Fiduciaries},
  author  = {Tuch, Andrew F.},
  journal = {Washington University Law Review},
  volume  = {98},
  number  = {6},
  pages   = {1897--1937},
  year    = {2021}
}

@inproceedings{kwon_etal_2024,
  title     = {Are {LLM}s Effective Negotiators? Systematic Evaluation of the Multifaceted Capabilities of {LLM}s in Negotiation Dialogues},
  author    = {Kwon, Deuksin and Weiss, Emily and Kulshrestha, Tara and Chawla, Kushal and Lucas, Gale M. and Gratch, Jonathan},
  booktitle = {Findings of the Association for Computational Linguistics: EMNLP 2024},
  pages     = {5391--5413},
  year      = {2024}
}

@inproceedings{bianchi_etal_2024,
  title     = {How Well Can {LLM}s Negotiate? {NegotiationArena} Platform and Analysis},
  author    = {Bianchi, Federico and Chia, Patrick John and Yuksekgonul, Mert and Tagliabue, Jacopo and Jurafsky, Dan and Zou, James},
  booktitle = {Proceedings of the 41st International Conference on Machine Learning},
  series    = {Proceedings of Machine Learning Research},
  volume    = {235},
  pages     = {3935--3951},
  publisher = {PMLR},
  year      = {2024}
}

@inproceedings{hua_etal_2024,
  title     = {Assistive Large Language Model Agents for Socially-Aware Negotiation Dialogues},
  author    = {Hua, Yuncheng and Qu, Lizhen and Haffari, Gholamreza},
  booktitle = {Findings of the Association for Computational Linguistics: EMNLP 2024},
  pages     = {8047--8074},
  year      = {2024}
}

@inproceedings{abdelnabi_etal_2024,
  title     = {Cooperation, Competition, and Maliciousness: {LLM}-Stakeholders Interactive Negotiation},
  author    = {Abdelnabi, Sahar and Gomaa, Amr and Sivaprasad, Sarath and Sch{\"o}nherr, Lea and Fritz, Mario},
  booktitle = {Advances in Neural Information Processing Systems, Datasets and Benchmarks Track},
  year      = {2024}
}

@article{raiffa_1985,
  title   = {Post-Settlement Settlements},
  author  = {Raiffa, Howard},
  journal = {Negotiation Journal},
  volume  = {1},
  number  = {1},
  pages   = {9--12},
  year    = {1985},
  doi     = {10.1111/j.1571-9979.1985.tb00286.x}
}

@article{ehtamo_etal_1999,
  title   = {Generating {P}areto Solutions in a Two-Party Setting: Constraint Proposal Methods},
  author  = {Ehtamo, Harri and H{\"a}m{\"a}l{\"a}inen, Raimo P. and Heiskanen, Pirja and Teich, Jeffrey E. and Verkama, Markku and Zionts, Stanley},
  journal = {Management Science},
  volume  = {45},
  number  = {12},
  pages   = {1697--1709},
  year    = {1999},
  doi     = {10.1287/mnsc.45.12.1697}
}

@article{rangaswamy_shell_1997,
  title   = {Using Computers to Realize Joint Gains in Negotiations: Toward an Electronic Bargaining Table},
  author  = {Rangaswamy, Arvind and Shell, G. Richard},
  journal = {Management Science},
  volume  = {43},
  number  = {8},
  pages   = {1147--1163},
  year    = {1997},
  doi     = {10.1287/mnsc.43.8.1147}
}

@inproceedings{schneider_etal_2024,
  title     = {Negotiating with {LLM}s: Prompt Hacks, Skill Gaps, and Reasoning Deficits},
  author    = {Schneider, Johannes and Haag, Steffi and Chandra Kruse, Leona},
  booktitle = {Computer-Human Interaction Research and Applications (CHIRA 2024)},
  series    = {Communications in Computer and Information Science},
  volume    = {2371},
  publisher = {Springer},
  pages     = {238--259},
  year      = {2024},
  doi       = {10.1007/978-3-031-83845-3_15}
}

@inproceedings{qian_etal_2026,
  title     = {Strategic Tradeoffs Between Humans and {AI} in Multi-Agent Bargaining},
  author    = {Qian, Crystal and Zhu, Kehang and Horton, John J. and Manning, Benjamin S. and Tsai, Vivian and Wexler, James and Thain, Nithum},
  booktitle = {Proceedings of the 31st International Conference on Intelligent User Interfaces (IUI '26)},
  note      = {arXiv:2509.09071},
  year      = {2026},
  doi       = {10.1145/3742413.3789078}
}

@article{fu_etal_2023,
  title   = {Improving Language Model Negotiation with Self-Play and In-Context Learning from {AI} Feedback},
  author  = {Fu, Yao and Peng, Hao and Khot, Tushar and Lapata, Mirella},
  journal = {arXiv preprint arXiv:2305.10142},
  year    = {2023}
}

@inproceedings{he_etal_2018,
  title     = {Decoupling Strategy and Generation in Negotiation Dialogues},
  author    = {He, He and Chen, Derek and Balakrishnan, Anusha and Liang, Percy},
  booktitle = {Proceedings of the 2018 Conference on Empirical Methods in Natural Language Processing (EMNLP)},
  pages     = {2333--2343},
  year      = {2018}
}

@article{choose_your_agent_2026,
  title   = {Choose Your Agent: Tradeoffs in Adopting {AI} Advisors, Coaches, and Delegates in Multi-Party Negotiation},
  author  = {Zhu, Kehang and Thain, Nithum and Tsai, Vivian and Wexler, James and Qian, Crystal},
  journal = {arXiv preprint arXiv:2602.12089},
  year    = {2026}
}

@article{mit_coaching_2025,
  title   = {Negotiation Backtable Bots: Using {GenAI} to Improve Multiparty Negotiation Instruction},
  author  = {Dinnar, Samuel and Susskind, Lawrence and Sibanda, Leroy and Olaleye, Ololade},
  journal = {Negotiation Journal},
  volume  = {41},
  pages   = {19--65},
  doi     = {10.1162/ngtn_a_00015},
  year    = {2025}
}

@techreport{eidenmueller_2026,
  author      = {Eidenmueller, Horst G. M.},
  title       = {From Contract Embellisher to Negotiation Machine: The Use of Artificial Intelligence in Contracting},
  institution = {SSRN},
  type        = {Working Paper},
  number      = {7682042},
  year        = {2026},
  month       = nov,
  doi         = {10.2139/ssrn.7682042},
  url         = {https://ssrn.com/abstract=7682042}
}

\appendix
\counterwithin*{figure}{section}
\counterwithin*{table}{section}
\renewcommand{\thefigure}{\thesection\arabic{figure}}
\renewcommand{\thetable}{\thesection\arabic{table}}

\clearpage

\section{Detailed Information on Experimental Design}
\label{app:procedure}

\subsection{Prolific Study Collection}

Participants were recruited through a Prolific study collection containing two arms, one per condition. Prolific randomly allocated each entering participant to exactly one arm. Within each condition, participants entered a multi-user lobby and were paired into two-person dyads in the order that they arrived. Study materials instructed each participant that their partner was another Prolific participant, and the advertised study duration was 20 minutes.

Participants were compensated with a \$3 base payment plus a performance bonus of up to \$3.

\subsection{Study Instructions}
\label{app:instructions}

Before the negotiation, all participants saw the following instructions, presented as sequential screens:

\begin{quote}
\underline{You will be paid based on your individual performance.}

You will earn a base payment of \$3.

You can earn a bonus payment of up to \$3 based on \underline{how good of a final agreement} you end up with. This determination is made based on the full agreement, including \underline{price} and \underline{other terms}.

You \underline{aren't competing with your partner for this bonus}. If you end up with a final agreement that is very good for both you and your partner, you can both independently earn a bonus.

If you do not reach an agreement with your partner, you are not eligible for a bonus.
\end{quote}

\begin{quote}
\underline{Your bonus will be calculated based on your final agreement ONLY.}

Your answers to the survey questions at the end will \underline{NOT} be used to calculate your bonus.

Therefore, please answer the survey questions candidly.
\end{quote}

Participants in the Treatment condition additionally saw:

\begin{quote}
\underline{After your negotiation, you will chat with an AI agent.}

We've designed an AI system to try to find win-win improvements to agreements.

The AI system will seek to understand what you want and what your partner wants.

The AI system will think about your negotiation and craft a new proposal, which you can choose to adopt or reject.
\end{quote}

\subsection{Exit Survey}
\label{app:survey}

The exit survey first asked participants to rate five statements on a seven-point scale (1 = Strongly disagree, 7 = Strongly agree):

\begin{itemize}
\item I am satisfied with the final agreement.
\item I think the other side is satisfied with the final agreement.
\item The final agreement feels fair to both sides.
\item The final agreement is clear, unambiguous, and easy to implement.
\item Overall, I feel satisfied with the process of reaching the final agreement.
\end{itemize}

The survey then collected demographic information:

\begin{itemize}
\item In the past 30 days, how often have you used AI tools? (e.g., ChatGPT, Gemini, Claude) --- Several times a day / About once a day / Several times a week / About once a week / Less than weekly / Never
\item How would you describe your experience with negotiations? (e.g., job offers, contracts, real estate) --- None / Beginner / Intermediate / Advanced / Expert
\item Which age range do you fall into? --- 18--24 / 25--34 / 35--44 / 45--54 / 55--64 / 65 or older
\item What is your race or ethnicity? (Select all that apply) --- American Indian or Alaska Native / Asian / Black or African American / Hispanic or Latino / Middle Eastern or North African / Native Hawaiian or Pacific Islander / White / Another race or ethnicity / Prefer not to answer
\item What is your gender identity? --- Man / Woman / Non-binary / Prefer not to answer
\item Which best describes your political affiliation? --- Democratic Party / Republican Party / Independent / Another party / Prefer not to answer
\item What was your total household income before taxes in the last 12 months? --- Less than \$25,000 / \$25,000--\$49,999 / \$50,000--\$74,999 / \$75,000--\$99,999 / \$100,000--\$149,999 / \$150,000--\$199,999 / \$200,000 or more / Prefer not to answer
\end{itemize}

\subsection{Negotiation Interface}

\begin{figure}[H]
\centering
\includegraphics[width=0.94\textwidth]{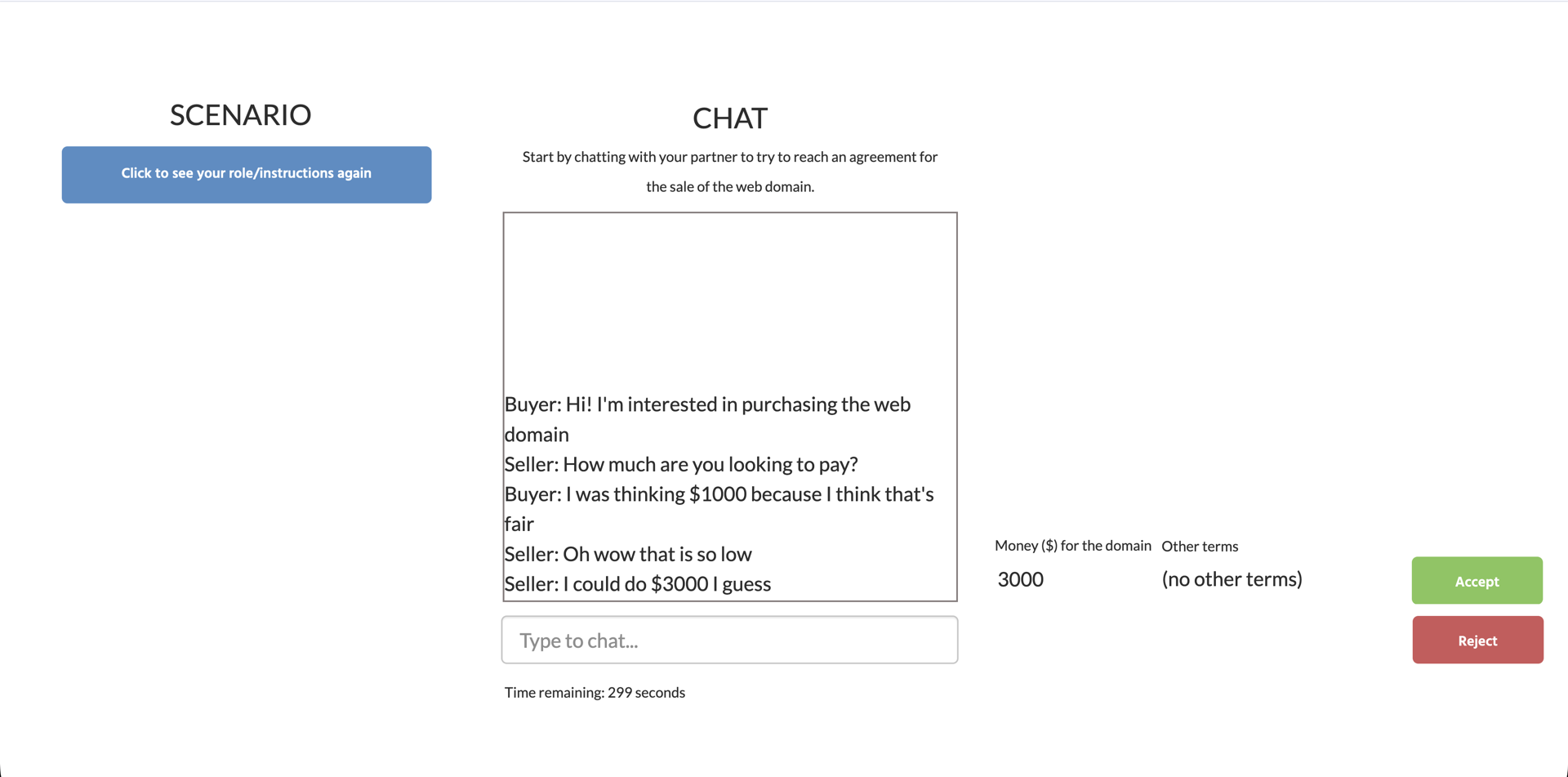}
\caption{The negotiation interface. Parties exchange free-form chat messages (center) and may compose a binding offer at any time by entering a price and optional non-price terms; a received offer (right) can be accepted, forming the initial agreement, or rejected. The scenario button (left) redisplays the participant's role instructions.}
\label{fig:interface-offer}
\end{figure}

\begin{figure}[H]
\centering
\includegraphics[width=0.94\textwidth]{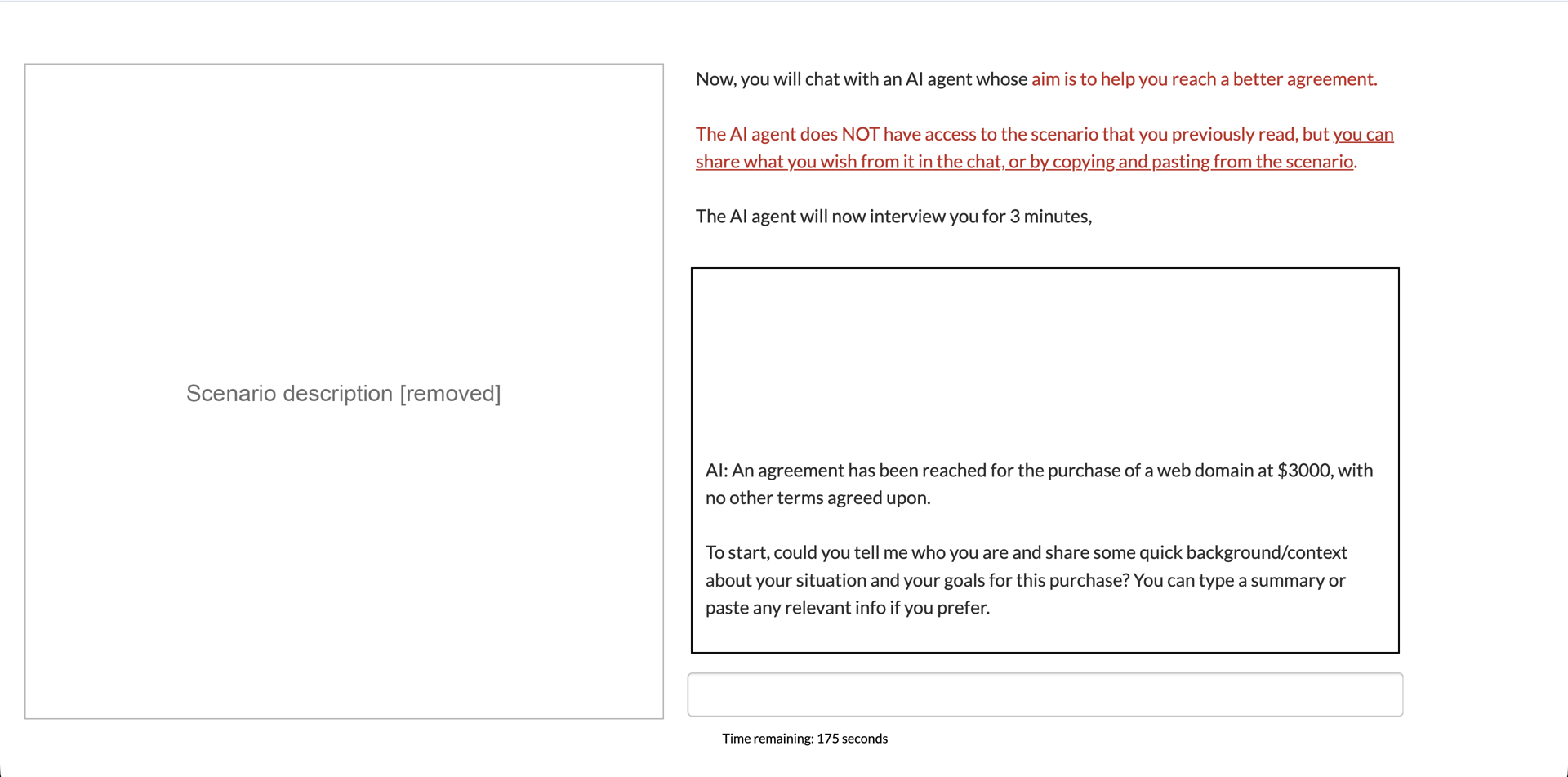}
\caption{The Information Fiduciary interview (Treatment condition). Each party is interviewed separately and privately; the agent has no access to the participant's role instructions (left panel, redacted) unless the participant chooses to share information from them.}
\label{fig:interface-fiduciary}
\end{figure}

\begin{figure}[H]
\centering
\includegraphics[width=0.94\textwidth]{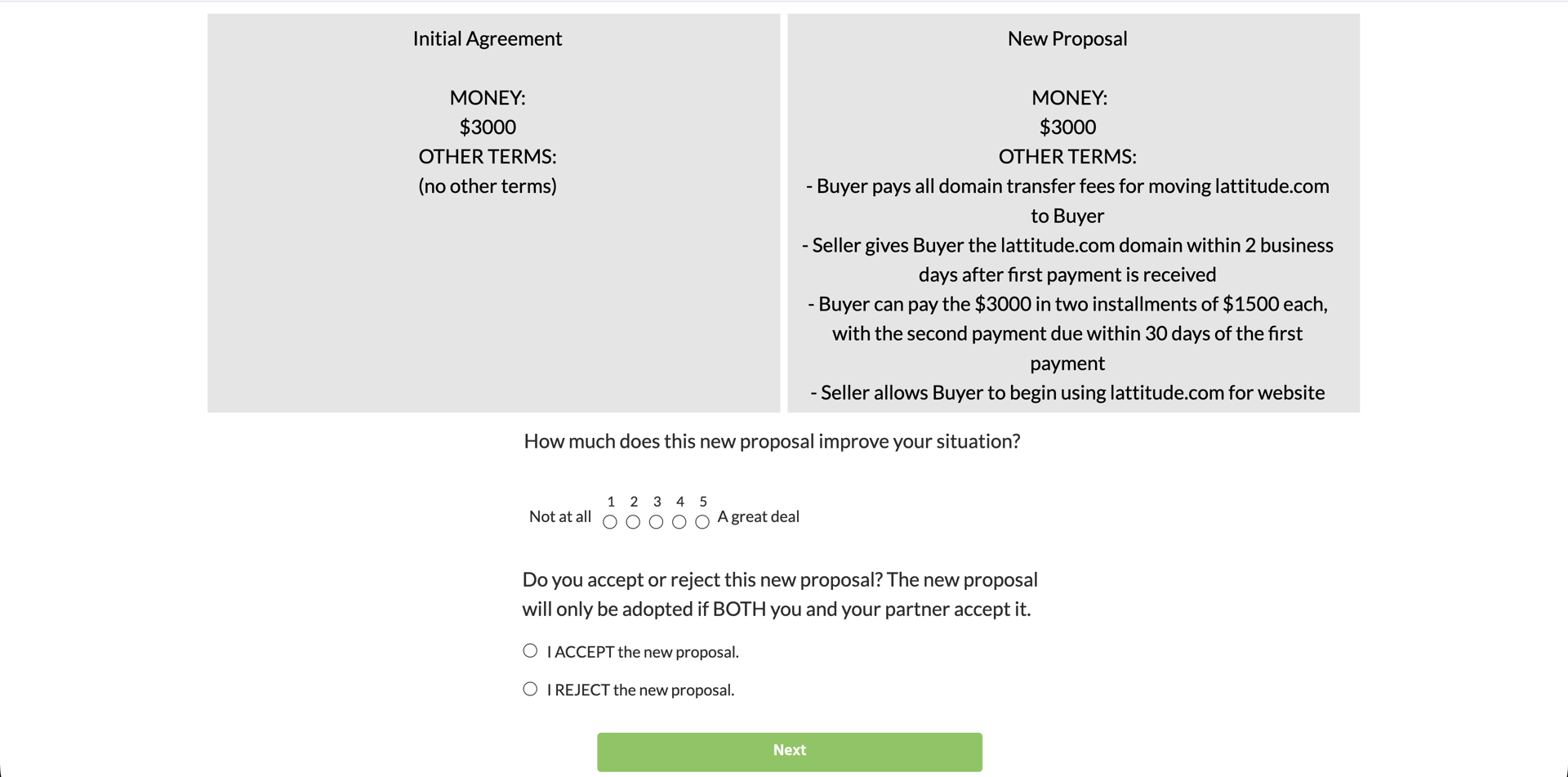}
\caption{The proposal review screen (Treatment condition). The initial agreement and the agent's proposed replacement are shown side by side; the replacement takes effect only if both parties independently accept it.}
\label{fig:interface-proposal}
\end{figure}

Dyads negotiated through a three-pane interface: a scenario button that redisplayed the participant's role instructions, a free-form text chat, and a structured offer panel. Either party could enter a price and optional non-price terms into the offer panel and submit it as a binding offer; the counterparty could accept, forming the initial agreement and ending the negotiation, or reject. Figures~\ref{fig:interface-offer}--\ref{fig:interface-proposal} show the participant-facing screens from a researcher test session: the negotiation interface, the confidential interview with the LLM Information Fiduciary, and the review screen on which each party privately accepted or rejected the proposed replacement agreement. The black box covers the participant's confidential role instructions, which are copyrighted case material \citep{manwaring_lattitude_2003}.

\section{Additional Results}
\label{app:results-figures}
\setcounter{figure}{0}
\setcounter{table}{0}

\subsection{Robustness to Within-Dyad Dependence}
\label{app:robustness}

\begin{table}[H]
\centering
\caption{Robustness of survey outcomes to within-dyad dependence. Each row reports an OLS regression of the outcome on a Treatment indicator with cluster-robust (CR2) standard errors clustered by dyad and Satterthwaite degrees of freedom (161 clusters). Secondary-outcome $p$ values are adjusted with the Benjamini--Hochberg procedure.}
\label{tab:robustness}
\resizebox{\textwidth}{!}{%
\begin{tabular}{lcccc}
\hline
Outcome & Mean difference [95\% CI] & CR2 SE & $p$ & BH-adjusted $p$ \\
\hline
Satisfied with the final agreement & 0.74 [0.39, 1.10] & 0.18 & $<.001$ & -- \\
Agreement feels fair to both sides & 0.71 [0.31, 1.11] & 0.20 & $<.001$ & .001 \\
Satisfied with the negotiation process & 0.64 [0.28, 1.01] & 0.18 & $<.001$ & .001 \\
Think the other side is satisfied & 0.35 [0.07, 0.63] & 0.14 & .014 & .018 \\
Agreement is clear and implementable & 0.04 [$-$0.24, 0.32] & 0.14 & .784 & .784 \\
\hline
\end{tabular}%
}
\end{table}

\subsection{Proposal Acceptance by Gender}
\label{app:gender}

As an exploratory analysis, we compared acceptance of the LLM-generated proposal by gender among Treatment participants whose dyad reached an initial agreement. Acceptance rates were 93.0\% for men (66 of 71) and 91.8\% for women (67 of 73), with no significant difference detected (Fisher's exact test, $p = 1.00$; odds ratio $= 1.18$, 95\% CI $[0.29, 5.14]$). Participants identifying as non-binary ($n = 3$) or declining to report gender ($n = 4$) were excluded from this analysis.

\subsection{Interview-Derived Preference Items}
\label{app:preferences}
\begin{table}[H]
\centering
\footnotesize
\caption{Interview-derived preference items from Treatment information-collection interviews ($n=152$), ranked by endorsement rate. For each interview and item the stance was coded Yes, No, Mixed/Depends, or Unclear by LLM. Endorsement is Yes\,$\div$\,coded, where coded excludes Unclear and not-mentioned. Mentioned refers to the number of interviews referencing the item.}
\label{tab:preferences}
\begin{tabular}{r p{0.46\textwidth} c c c}
\hline
Rank & Preference item & Yes / coded & Endorsement & Mentioned \\
\hline
1 & Clean reputation and technical assurances & 7 / 7 & 100.0\% & 7 \\
2 & Ownership and legal assurances & 16 / 18 & 88.9\% & 19 \\
3 & Documentation and contract & 17 / 21 & 81.0\% & 22 \\
4 & Privacy and confidentiality & 11 / 14 & 78.6\% & 14 \\
5 & Transaction fee responsibility & 39 / 50 & 78.0\% & 52 \\
6 & Escrow or payment security & 8 / 11 & 72.7\% & 11 \\
7 & Payment method & 30 / 42 & 71.4\% & 43 \\
8 & Payment speed & 36 / 51 & 70.6\% & 52 \\
9 & Transfer timing and closing deadline & 53 / 80 & 66.3\% & 81 \\
10 & Updates and confirmations & 7 / 12 & 58.3\% & 12 \\
11 & Payment structure & 48 / 83 & 57.8\% & 83 \\
12 & Brand exclusivity and use restrictions & 13 / 23 & 56.5\% & 23 \\
13 & Domain history and metrics & 6 / 11 & 54.5\% & 12 \\
14 & Transfer responsibility and assistance & 49 / 94 & 52.1\% & 95 \\
15 & Future or contingent compensation & 14 / 28 & 50.0\% & 28 \\
16 & Post-transfer support & 12 / 24 & 50.0\% & 24 \\
17 & Contingencies, refunds, and remedies & 10 / 20 & 50.0\% & 20 \\
18 & Upfront price adjustment & 13 / 27 & 48.1\% & 27 \\
19 & Transition access and routing & 7 / 16 & 43.8\% & 16 \\
20 & Related assets bundle & 8 / 21 & 38.1\% & 22 \\
21 & Future purchase or buyback rights & 7 / 23 & 30.4\% & 23 \\
22 & Public recognition or promotion & 29 / 105 & 27.6\% & 109 \\
23 & Non-cash perks and resources & 4 / 17 & 23.5\% & 17 \\
24 & Professional opportunities and relationships & 13 / 57 & 22.8\% & 57 \\
\hline
\end{tabular}
\end{table}

\subsection{Supplementary Figures}
\label{app:figures}

\begin{figure}[H]
\centering
\includegraphics[width=0.94\textwidth]{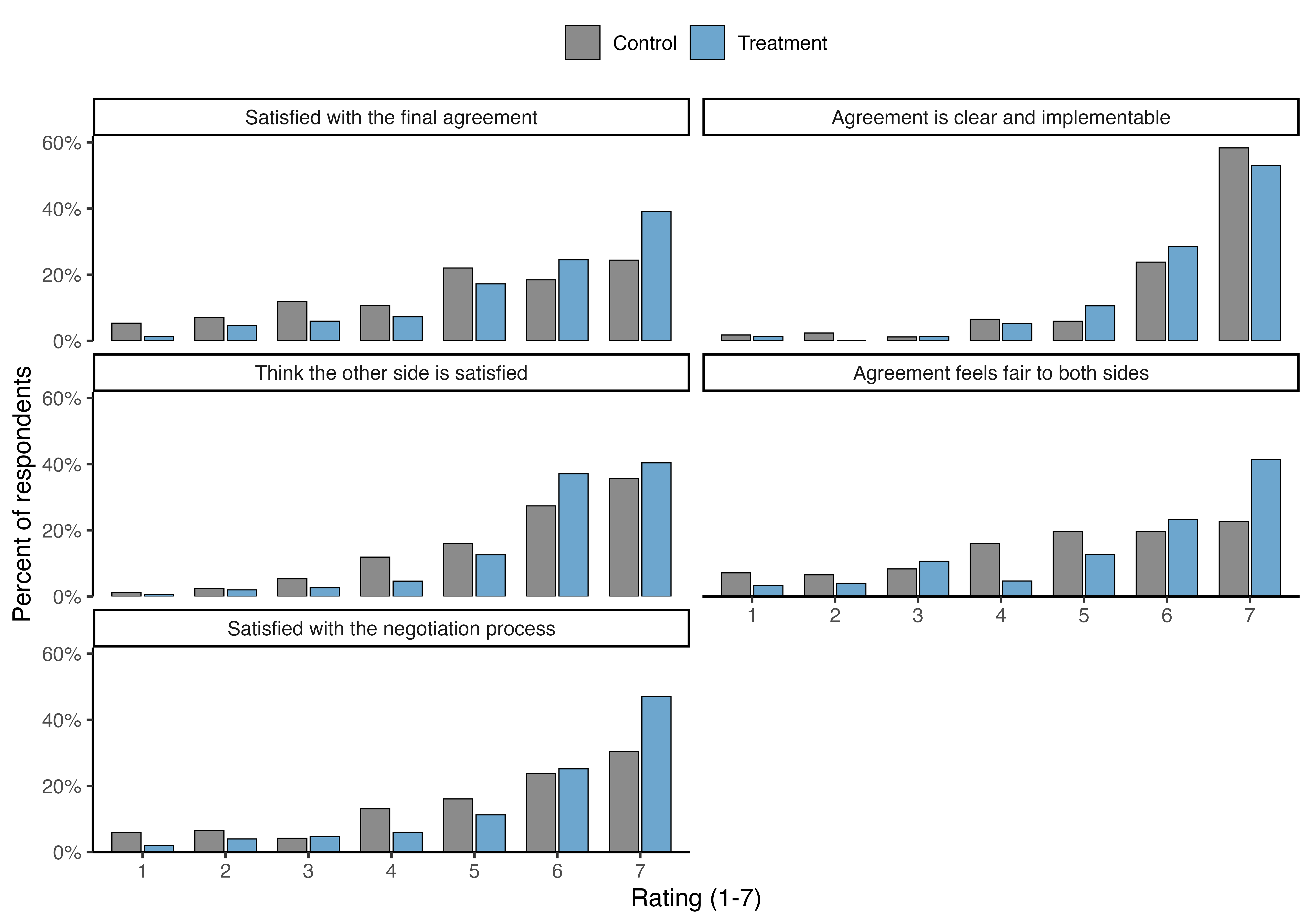}
\caption{Response distributions for each survey outcome by condition. Bars show the within-condition percentage of responses at each Likert rating.}
\label{fig:distributions}
\end{figure}

\begin{figure}[H]
\centering
\includegraphics[width=0.86\textwidth]{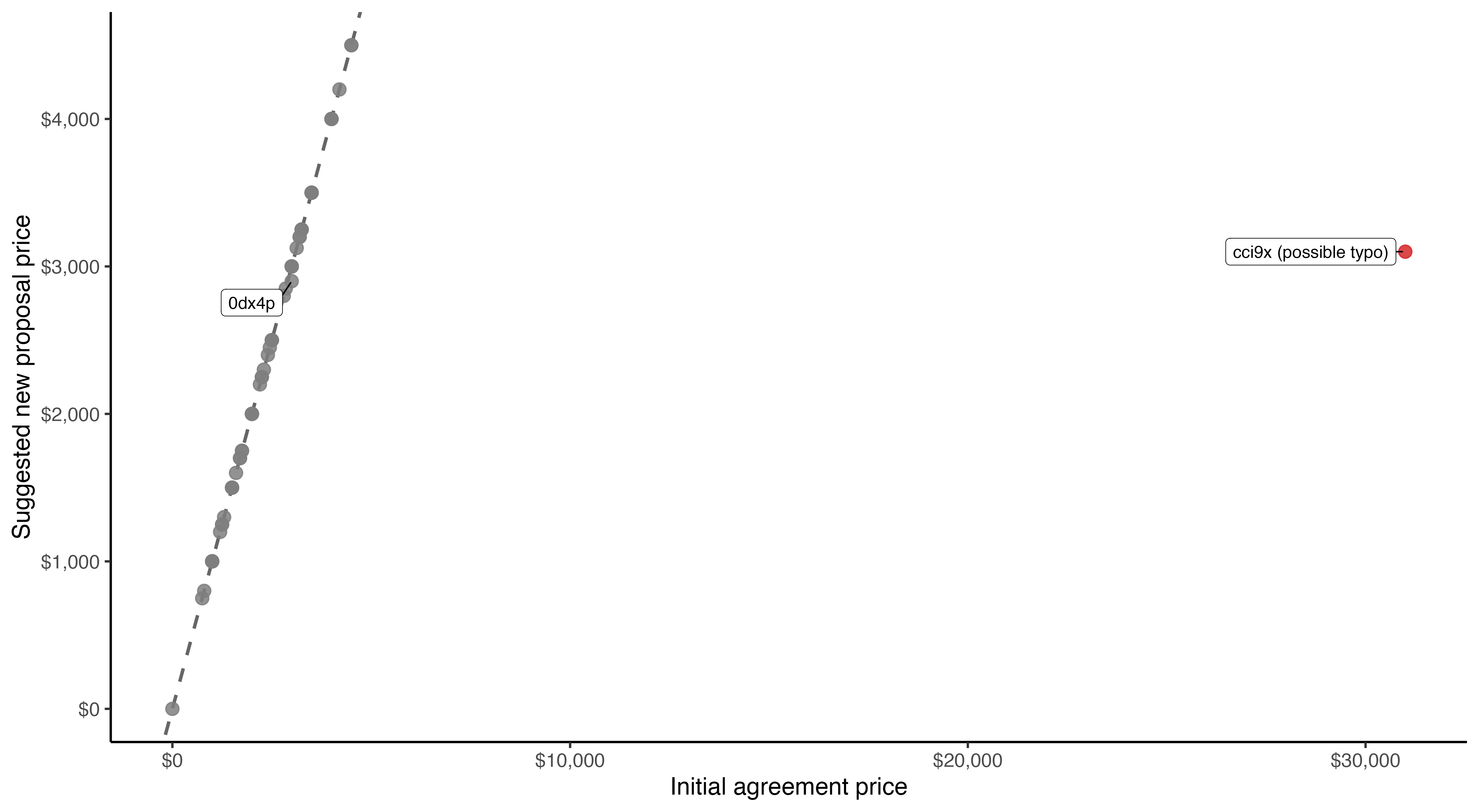}
\caption{Initial agreement price versus suggested new proposal price among Treatment dyads that reached agreement. Each point is a dyad; the dashed line indicates equality.}
\label{fig:price}
\end{figure}

\begin{figure}[H]
\centering
\includegraphics[width=0.86\textwidth]{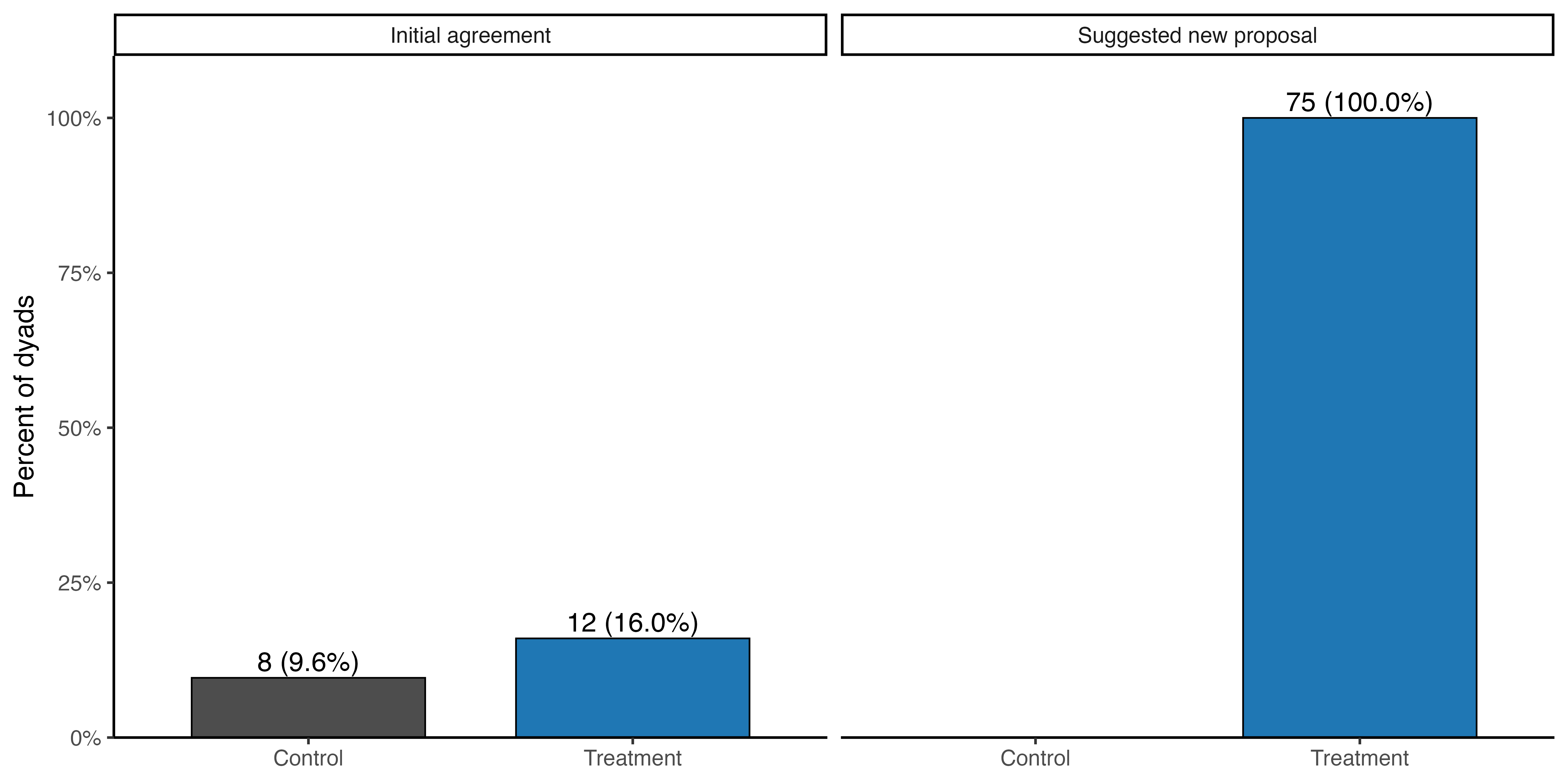}
\caption{Presence of any non-price ``other'' terms among dyads that reached agreement. Bars show the percentage of dyads with at least one other-term clause in initial agreements and in Treatment proposals.}
\label{fig:presence}
\end{figure}

\begin{figure}[H]
\centering
\includegraphics[width=0.86\textwidth]{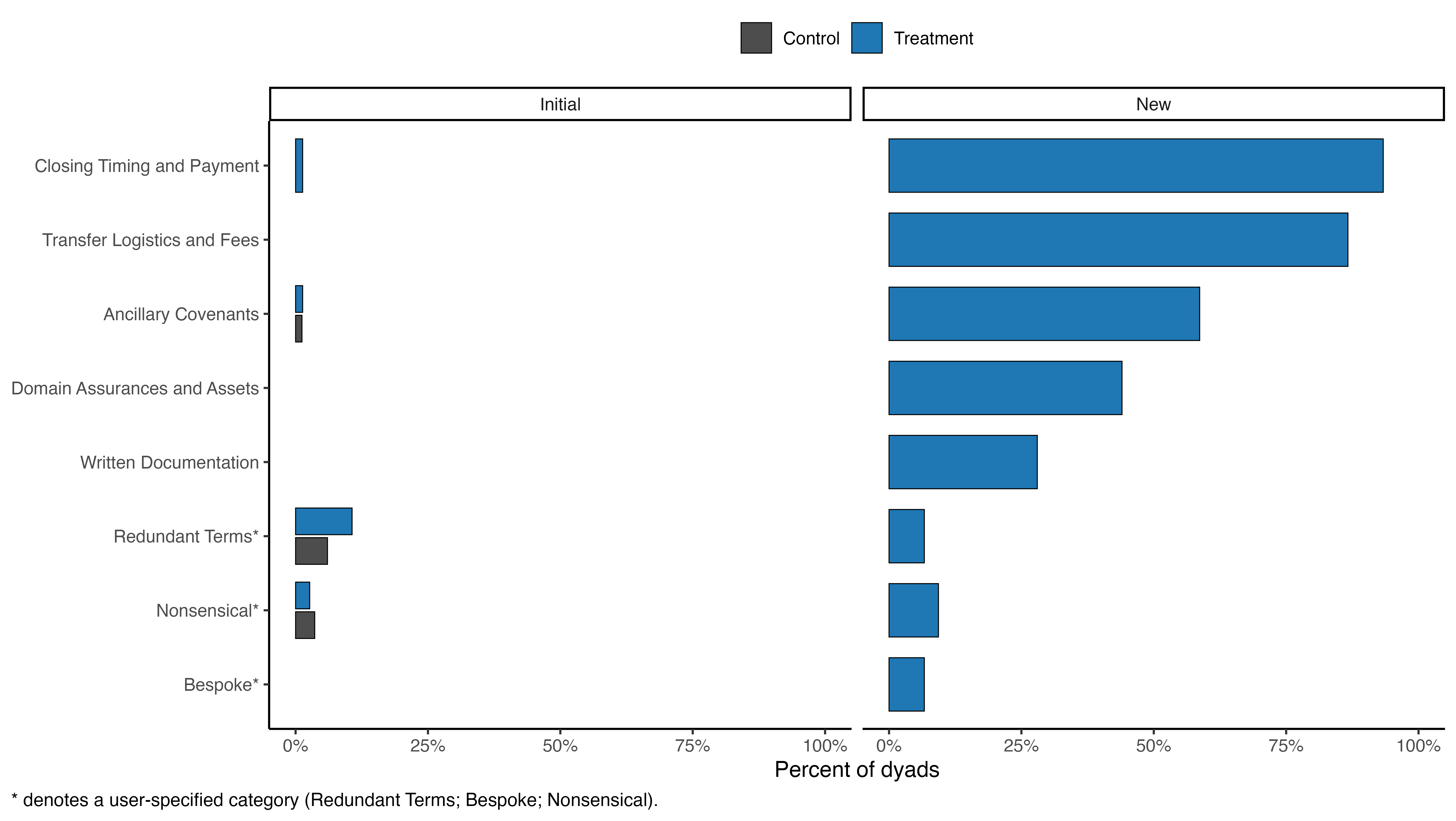}
\caption{Prevalence of other-term categories among dyads that reached agreement. Bars show the percentage of dyads with at least one clause in each category for initial agreements and suggested new proposals.}
\label{fig:categories}
\end{figure}

\clearpage
\end{document}